\documentclass[twocolumn,resetfootnote]{aastex701}

\defcitealias{Zou2023ApJ...950..136Z}{Z23}
\usepackage{amsmath}
\usepackage{CJK}
\usepackage{placeins}

\begin{document}

\title{Deep Multiwavelength Constraints on the Host Galaxies and Black-Hole Growth of Dwarf AGNs}

\begin{CJK*}{UTF8}{gbsn}

\author[orcid=0000-0003-0747-1780]{Wei Leong Tee}
\affiliation{Department of Astronomy and Astrophysics, The Pennsylvania State University, 525 Davey Lab, University Park, PA 16802, USA}
\affiliation{Institute for Gravitation and the Cosmos, The Pennsylvania State University, University Park, PA 16802, USA}
\email[show]{wmt5159@psu.edu}  

\author[orcid=0000-0002-0167-2453]{W. N. Brandt} 
\affiliation{Department of Astronomy and Astrophysics, The Pennsylvania State University, 525 Davey Lab, University Park, PA 16802, USA}
\affiliation{Institute for Gravitation and the Cosmos, The Pennsylvania State University, University Park, PA 16802, USA}
\affiliation{Department of Physics, The Pennsylvania State University, 104 Davey Laboratory, University Park, PA 16802, USA}
\email{wnbrandt@gmail.com}

\author[orcid=0000-0002-6990-9058,gname=Zhibo,sname=Yu]{Zhibo Yu (喻知博)}
\affiliation{Department of Astronomy and Astrophysics, The Pennsylvania State University, 525 Davey Lab, University Park, PA 16802, USA}
\affiliation{Institute for Gravitation and the Cosmos, The Pennsylvania State University, University Park, PA 16802, USA}
\email{zvy5225@psu.edu}

\author[orcid=0000-0002-4436-6923,gname=Fan,sname=Zou]{Fan Zou}
\affiliation{Department of Astronomy, University of Michigan, 1085 S University, Ann Arbor, MI 48109, USA}
\email{fanzou01@gmail.com}

\author[orcid=0009-0009-8209-4613]{Bojun Zhang}
\affiliation{Department of Astronomy and Astrophysics, The Pennsylvania State University, 525 Davey Lab, University Park, PA 16802, USA}
\affiliation{Institute for Gravitation and the Cosmos, The Pennsylvania State University, University Park, PA 16802, USA}
\email{bjz5234@psu.edu}
\begin{abstract}
We present a multiwavelength population study of 56 X-ray-detected dwarf galaxies ($M_* < 3 \times 10^9 M_\odot$), comprising 31 active galactic nuclei (AGNs) and 25 galaxies (not satisfying AGN classification) in the Chandra Deep Fields North and South. 
Utilizing ultra-deep Chandra observations together with high-resolution JWST and HST imaging, we investigate the demographics, host-galaxy properties, environments, and cosmic evolution of accreting black holes in low-mass galaxies out to $z\approx4$. 
We find no statistically significant differences in either the local environments or non-parametric morphologies of X-ray AGNs relative to mass- and redshift-matched inactive dwarf galaxies. These results suggest that AGN activity in dwarf galaxies does not generally require ongoing major mergers or strong large-scale disturbances. 
Using a single power-law model for the specific accretion-rate distribution, we infer a positive cosmic evolution of dwarf-galaxy black-hole growth. Both the observed AGN fraction and the long-term average black-hole accretion rate increase toward higher redshift and show a positive stellar-mass dependence. 
Many dwarf AGNs exhibit weak optical and near-infrared AGN signatures, consistent with moderate accretion rates and, at high redshift, potentially enhanced obscuration. The small number of detected systems at $z>2$ limits definitive conclusions, but their hard X-ray colors suggest that obscuration may play an increasingly important role in the dwarf AGN population in the early universe. 
Our results support a picture in which black-hole growth in dwarf galaxies proceeds through largely stochastic or secular processes while gradually increasing in prevalence toward cosmic noon, providing new empirical constraints on the low-mass end of black hole-galaxy coevolution.
\end{abstract}


\section{Introduction} 
The origins of supermassive black holes (SMBHs) remain a fundamental open question, especially following the discovery of billion-solar-mass quasars at $z>6$ \citep[e.g.,][]{Fan2023ARA&A..61..373F}. 
To explain the rapid assembly of these systems within the first billion years of cosmic history, theoretical models propose a range of early-universe seeding channels \citep[e.g.,][]{Inayoshi2020ARA&A..58...27I}. 
These scenarios are commonly grouped into ``light" seeds, with characteristic masses of $\sim 100~M_\odot$ and originating from Population III stellar remnants, and ``heavy" seeds, with characteristic masses of $\sim 10^{4}$--$10^{5}~M_\odot$ forming through the direct collapse of pristine gas clouds \citep[e.g.,][]{Volonteri2012Sci...337..544V,Natarajan2014GReGr..46.1702N,Inayoshi2020ARA&A..58...27I}. 
Because dwarf galaxies are expected to have experienced comparatively limited merger-driven growth, their central black holes (BHs) may preserve relatively clean signatures of early black-hole formation. 
Dwarf galaxies therefore provide a critical laboratory for connecting present-day BH demographics to early-universe seeding models \citep[e.g.,][]{Greene2020ARA&A..58..257G,Reines2022NatAs...6...26R}. Distinguishing between these competing seeding scenarios requires measuring the BH occupation fraction, $f_{\mathrm{occ}}$, in the low-mass galaxy regime. 
Heavy-seed models generally predict a strong decline in occupation fraction toward dwarf-galaxy masses, whereas light-seed models predict a substantially higher, and possibly nearly ubiquitous, BH occupation fraction \citep[e.g.,][]{Ricarte2018MNRAS.481.3278R,Burke2025ApJ...978...77B,Zou2025ApJ...992..176Z}.

Historically, efforts to constrain BH demographics in low-mass galaxies have relied heavily on local ($z\approx0-1$) dwarf-galaxy populations \citep[e.g.,][]{Reines2015ApJ...813...82R,Mezcua2018MNRAS.478.2576M,Birchall2020MNRAS.492.2268B,Zou2023ApJ...950..136Z,Bykov2024MNRAS.527.1962B}. 
However, these local constraints provide only a limited view of the broader evolutionary picture because they are fundamentally limited by the cosmic decline of BH accretion. 
Instead, measuring the observed active galactic nucleus (AGN) fraction, $f_{\mathrm{AGN}}$, as a function of galaxy mass and cosmic time provides a direct observational constraint on the prevalence of accretion activity and the growth of low-mass BH populations. Such measurements are particularly valuable in dwarf galaxies, where AGN samples probe both the lowest-mass actively accreting BHs in the local universe and the progenitors of SMBHs at earlier cosmic epochs.

Multiwavelength AGN investigations are essential in this context, as no single observational tracer captures the full diversity of accretion states and obscuration levels \citep[e.g.,][]{Lyu2022ApJ...941..191L}. 
Combining X-ray, infrared, optical, and radio diagnostics enables a more complete census of AGN activity across a wide range of black-hole masses, accretion rates, and host-galaxy environments, particularly in the low-mass regime where individual selection techniques are strongly incomplete.
Extending these studies beyond the local universe is also critical because AGN activity rises strongly with redshift \citep[e.g.,][]{Aird2012ApJ...746...90A,Aird2018MNRAS.474.1225A,Yang2018MNRAS.475.1887Y,Zou2024ApJ...964..183Z}. 
As a result, the observable AGN fraction increases toward cosmic noon, allowing a larger fraction of the underlying BH population to be identified through accretion signatures. 
Characterizing the evolution of $f_{\mathrm{AGN}}$ and overall BH accretion in dwarf galaxies therefore provides an important empirical benchmark for understanding how low-mass BHs grow over cosmic time and how the earliest SMBH populations emerged.

While the abundance of AGNs constrains when and how frequently BH growth occurs, understanding the physical mechanisms responsible for that growth requires connecting AGN activity to host-galaxy properties. The environments, morphologies, and structural characteristics of AGN host galaxies provide important clues regarding the processes that trigger AGN activity. 
Measurements of these host-galaxy properties therefore complement demographic studies by revealing not only how often dwarf-galaxy BHs grow, but also the physical conditions under which that growth is triggered. 
In massive galaxies, particularly at high redshift, AGN activity is often associated with major hierarchical mergers or tidal interactions with massive neighbors, which can drive gas inflows and produce disturbed host morphologies \citep[e.g.,][]{Hopkins2006ApJS..163....1H,Triester2012ApJ...758L..39T,Bonaventura2025ApJ...978...74B}. 
However, it remains unclear whether such external, environment-driven interactions are prerequisites for triggering AGN activity in dwarf galaxies, or whether accretion is primarily governed by internal stochastic or secular processes \citep[e.g.,][]{Satyapal2014ApJ...784..113S,Martini2018MNRAS.476.2801M,Zhong2022ApJ...925..157Z,Wang2026ApJ..1003..172W}. 
Statistical comparisons of the environments and structural properties of active dwarf galaxies with matched inactive control samples therefore provide an independent way to test possible triggering mechanisms.

In this work, we combine the exceptional X-ray sensitivity of the Chandra Deep Fields (CDFs), consisting of the 2 Ms Chandra Deep Field-North (CDF-N) and the 7 Ms Chandra Deep Field-South (CDF-S), with deep, high-resolution multiwavelength imaging from the DAWN James Webb Space Telescope (JWST) Archive in the Great Observatories Origins Deep Survey (GOODS) fields. 
By combining JWST's infrared sensitivity with deep X-ray constraints, we can characterize X-ray-emitting dwarf galaxies to higher redshift and lower stellar mass than before. This enables us to construct a robust, host-galaxy mass-complete sample of X-ray-selected dwarf AGNs at $z\approx0-4$. Using this sample, we assess the structural and environmental drivers of AGN activity in dwarf host galaxies and quantify the dwarf AGN accretion properties. 

This paper is organized as follows. 
In Section 2, we describe the X-ray and multiwavelength data, source cross-matching, SED fitting, and dwarf AGN classification. 
In Section 3, we detail our approach to extracting morphology parameters and modeling the AGN specific accretion-rate distribution. 
In Section 4, we present our results regarding X-ray hardness ratio, host-galaxy environments, morphological properties, the cosmic evolution of the AGN fraction, and the long-term black-hole accretion rate. 
In the final section, we summarize this study and discuss future directions for improving and extending its methodology. 
Throughout this work, all quoted errors and uncertainties are reported at the $1\sigma$ ($68\%$) confidence level unless otherwise noted. 
We adopt a hard X-ray luminosity of $L_\mathrm{X}=L_\mathrm{X}$(2--10 keV) = 0.721 $L_\mathrm{X}$(0.5--7 keV), assuming an intrinsic power-law photon index of $\Gamma=1.8$.
We adopt a flat $\Lambda$CDM cosmology based on the \textit{Planck} 2018 results (\citealt{Planck2020A&A...641A...6P}), with $H_0=67.66~\mathrm{km~s^{-1}~Mpc^{-1}}$, $\Omega_{\mathrm{m}}=0.3111$, and $\Omega_{\Lambda}=0.6889$.

\section{Data and Sample}
\subsection{X-ray and Multiwavelength Data}
To construct the parent X-ray sample for this study, we utilize ultra-deep X-ray observations combined with comprehensive multiwavelength imaging from the DAWN JWST Archive \citep[DJA; e.g.,][]{Valentino2023ApJ...947...20V,brammer_2023_8370018} for the sub-regions within the GOODS fields that have JWST coverage. 
For the X-ray point-source (PS) data, we integrate the legacy catalogs from the 2 Ms Chandra Deep Field-North \citep[CDF-N;][]{Xue2016ApJS..224...15X} and the 7 Ms Chandra Deep Field-South \citep[CDF-S;][]{Luo2017ApJS..228....2L}.
\begin{figure*}[ht]
    \centering
    \includegraphics[width=\linewidth]{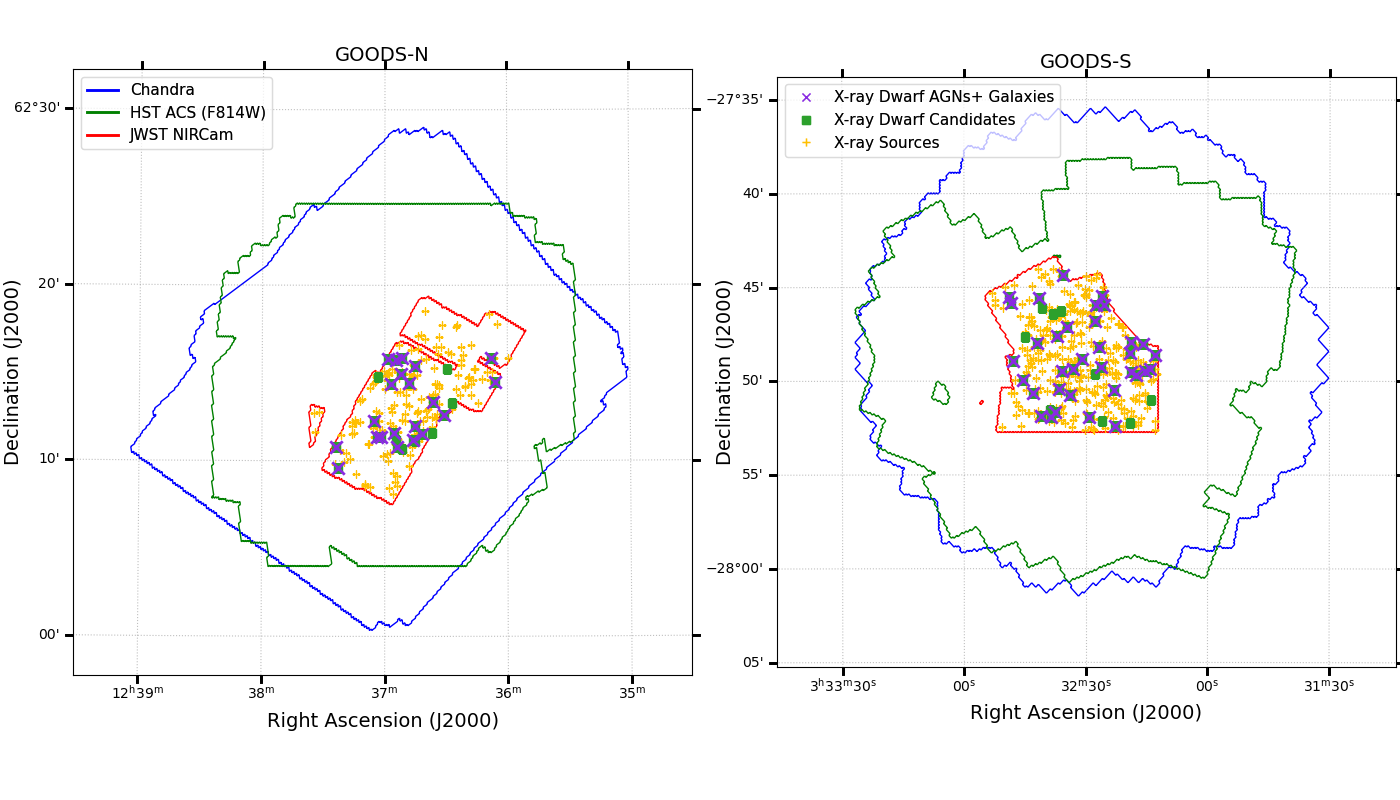}
    \caption{
    Survey footprints and spatial distributions of X-ray-detected sources in the GOODS-N (left) and GOODS-S (right) fields. Solid contours denote the observational coverage for Chandra (blue), HST ACS F814W (green), and JWST NIRCam (red; defined as the overlapping region between the F200W and F444W filters). 
    Symbols represent sources located within the JWST footprint, classified as follows: Chandra-detected X-ray sources (yellow pluses), X-ray dwarf candidates satisfying the mass threshold (green squares), and the final sample of X-ray dwarf AGNs and galaxies (purple crosses).
    }
    \label{fig:footprint}
\end{figure*}

For the optical and near-infrared (NIR) data, we employ the DJA's homogeneous, astrometrically matched (matched-WCS) reductions of publicly available imaging from both the Hubble Space Telescope (HST) and the James Webb Space Telescope (JWST). This combined dataset spans a wide wavelength range from the optical to the NIR, integrating deep observations from a multitude of legacy programs. 
These include the Cosmic Assembly Near-infrared Deep Extragalactic Legacy Survey \citep[CANDELS;][]{Grogin2011ApJS..197...35G,Koekemoer2011ApJS..197...36K}, 3D-HST \citep{Brammer2012ApJS..200...13B,Skelton2014ApJS..214...24S}, the JWST Advanced Deep Extragalactic Survey \citep[JADES;][]{Eisenstein2026ApJS..283....6E}, the First Reionization Epoch Spectroscopically Complete Observations \citep[FRESCO;][]{Oesch2023MNRAS.525.2864O}, and PANORAMIC \citep{Williams2025ApJ...979..140W}. 

As illustrated in Figure~\ref{fig:footprint}, our primary analysis is restricted to the well-studied regions of the Great Observatories Origins Deep Survey North (GOODS-N) and South (GOODS-S) fields that feature overlapping X-ray data, ancillary multiwavelength coverage, and JWST imaging (specifically the F200W and F444W bands to ensure homogeneous coverage).  
The effective area of the imaging footprint used in this work is $\sim57$ arcmin$^2$ in GOODS-N and $\sim60$ arcmin$^2$ in GOODS-S.

\subsection{Source Cross-Matching and Photometric Redshifts}
High-quality optical-to-NIR counterparts for the X-ray sources were previously compiled by \citet{Xue2016ApJS..224...15X} and \citet{Luo2017ApJS..228....2L}. 
The DJA JWST catalogs (specifically versions \texttt{gdn-grizli-v7.3} and \texttt{gds-grizli-v7.2}) include all sources detected within the JWST footprint \citep{Valentino2023ApJ...947...20V,brammer_2023_8370018}. 
These catalogs provide optical--NIR photometry, photometric redshifts ($z_\mathrm{phot}$), spectroscopic redshifts ($z_\mathrm{spec}$, when available), and derived galaxy physical properties, such as stellar mass ($M_\ast$), star-formation rate (SFR), and rest-frame \textit{UVJ} colors. 
The photometric redshifts and physical properties were derived using the spectral energy distribution (SED) fitting code \texttt{EAZY} \citep{Brammer2008ApJ...686.1503B}, which fits the observed photometry to galaxy templates to determine the maximum-likelihood redshift.\footnote{Galaxy-template-based photometric redshifts can be biased for AGNs whose optical--NIR emission is dominated by the active nucleus. For our dwarf AGN candidates, however, the optical--NIR emission is host-galaxy dominated, and the imaging does not reveal strong unresolved nuclear PS components, as discussed in Section~\ref{sec:nuclear_ps}. We therefore consider the \texttt{EAZY}-based photometric redshifts appropriate for this sample.} 
The vast majority of sources in GOODS-N and GOODS-S possess 8--20 and 11--29 high-quality bands (with signal-to-noise S/N $> 5$), respectively. 
This abundance of high-significance photometry ensures robust $z_\mathrm{phot}$ estimation across a wide range of redshifts. Finally, we cross-match the JWST catalogs with the optical--NIR counterparts of the X-ray sources using a matching radius of 0\farcs5.

\begin{figure*}[ht]
    \centering
    \includegraphics[width=\linewidth]{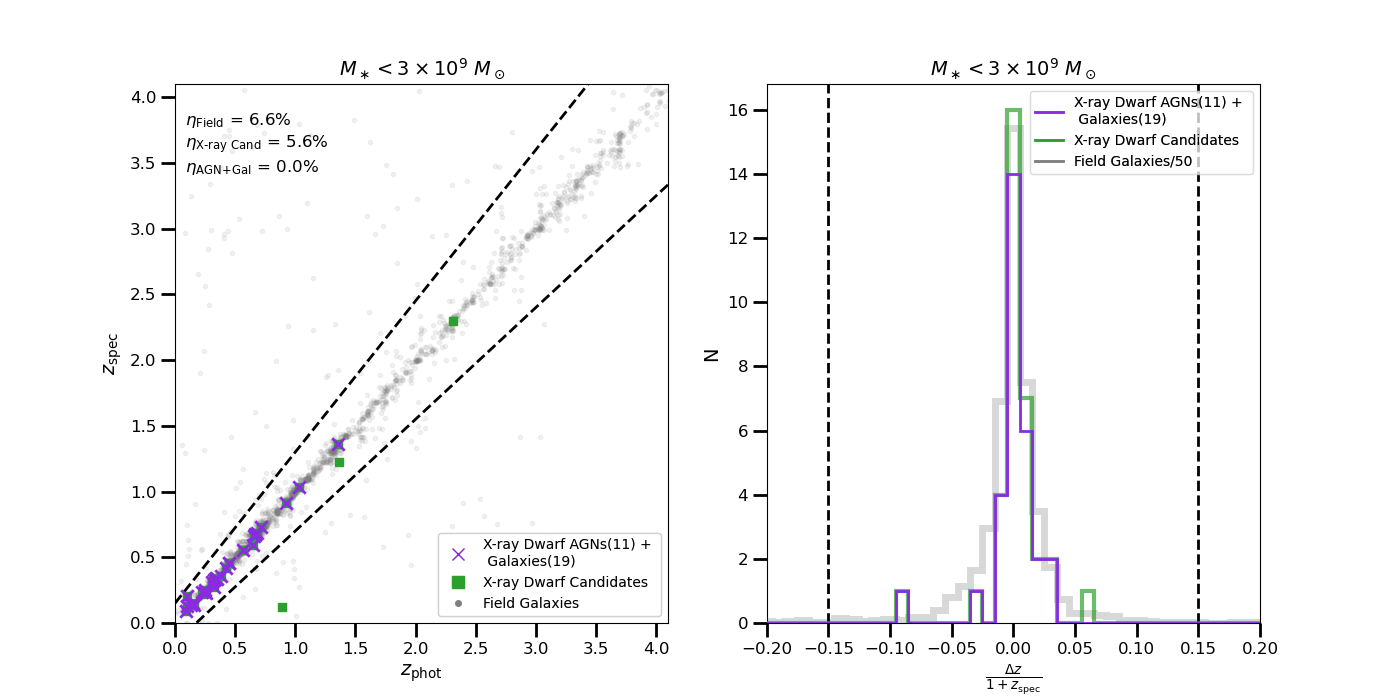}
    \caption{
    Photometric redshift reliability for the low-mass ($M_* < 3 \times 10^9\,M_\odot$) sample, comparing $z_\mathrm{spec}$ and $z_\mathrm{phot}$ from the Chandra+JWST multiwavelength catalog, shown in redshift space and as a  histogram. Green squares represent X-ray dwarf candidates satisfying the stellar mass threshold, while purple crosses denote the final sample of X-ray dwarf AGNs and galaxies; gray points indicate the underlying field galaxy population. 
    The left panel displays the direct comparison between photo-$z$ and available spec-$z$, whereas the right panel shows the corresponding distribution of residuals, $\Delta z/(1 + z_\mathrm{spec})$, with the field galaxy distribution scaled down by a factor of 50 to match the apparent distribution of the X-ray sources.
    Black dashed lines denote the $\pm0.15$ outlier threshold. Outlier rates ($\eta$) are provided in the upper-left corner of the left panel, demonstrating high photometric redshift reliability ($\eta \lesssim 10\%$) for both field galaxies and X-ray sources. 
    The apparently low outlier rate for the AGN and galaxy sample is primarily due to small-number statistics among the spectroscopically confirmed X-ray dwarf sources.
    }
    \label{fig:redshift_quality}
\end{figure*}

As highlighted by \citet[][hereafter \citetalias{Zou2023ApJ...950..136Z}]{Zou2023ApJ...950..136Z}, photometric-redshift quality is a critical systematic uncertainty in the selection of dwarf AGN, as poor $z_{\rm phot}$ constraints can severely bias the derived $M_\ast$ and SFR used to quantify host-galaxy properties. 
To assess the reliability of the $z_\mathrm{phot}$, we compare $z_\mathrm{phot}$ against available $z_\mathrm{spec}$ within the catalog. 
Among the final X-ray-detected dwarf sample, $\sim50\%$ of X-ray detected sources have reliable $z_\mathrm{spec}$. 
For the broader low-mass galaxy population with spectroscopic coverage, the resulting outlier fraction is $\lesssim$ 10\% (Figure~\ref{fig:redshift_quality}). 
Furthermore, we evaluate the photometric-redshift quality parameter defined by \citet{Brammer2008ApJ...686.1503B}, $\displaystyle Q_z=\frac{\chi^2}{N_\mathrm{filt}-3}\frac{z_\mathrm{up}^{99}-z_\mathrm{lo}^{99}}{p_{\Delta z=0.2}}$, where $\chi^2$ is the goodness-of-fit of the best-fitting template solution, $N_\mathrm{filt}$ is the number of photometric filters used in the fit, and $z_\mathrm{lo}^{99}$ and $z_\mathrm{up}^{99}$ are the lower and upper bounds of the central 99\% interval of the redshift posterior probability distribution. The quantity $p_{\Delta z=0.2}$ is the fraction of the total integrated redshift probability within $z_\mathrm{phot}\pm0.2$. Smaller $Q_z$ values indicate higher photometric-redshift quality. Among the more than 40,000 field dwarf galaxies, $64\%$ have $Q_z<1$ and $24\%$ have $Q_z>3$; the corresponding fractions for the 72 X-ray-detected dwarf candidates are $94\%$ and $3\%$, respectively.
These results support the reliability of the photometric redshifts used for the X-ray-detected dwarfs sample. 
Additionally, we visually inspected the photometric-redshift probability distributions of all candidates and confirmed that their redshift solutions exhibit well-defined, localized peaks rather than broad or multimodal distributions.
Nevertheless, because a substantial fraction of the X-ray AGNs still rely on photometric redshifts, redshift-dependent quantities such as $M_\ast$, SFR, and X-ray luminosity ($L_\mathrm{X}$) remain subject to the associated photo-$z$ uncertainties.

\subsection{SED Fitting and Dwarf-Galaxy Selection}\label{sec:sed_fitting_dwarf_selection}
Initial $M_\ast$ estimates are compiled from \citet{Santini2015ApJ...801...97S}, \citet{Barro2019ApJS..243...22B}, and \citet{Lyu2022ApJ...941..191L} where available. 
Additionally, \texttt{EAZY} provides a first-order mass estimation assuming that all measured fluxes are dominated by stellar emission from the host galaxy. 
In cases where the \texttt{EAZY} mass is utilized, we acknowledge that the presence of an AGN may introduce bias to the host galaxy mass estimate; we accept this initial contamination to maintain sample completeness before proceeding to rigorous SED fitting to explicitly decouple the AGN contribution.
\begin{figure*}[ht]
    \centering
    \includegraphics[width=\linewidth]{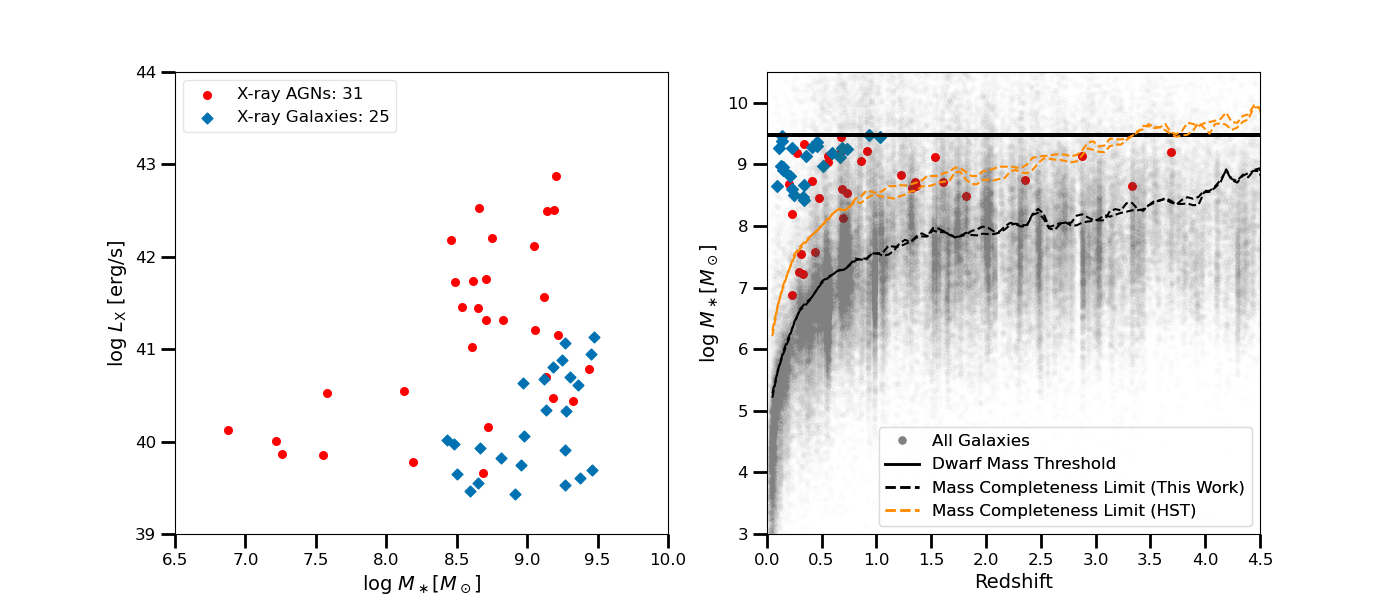}
    \caption{\textit{Left:} Distribution of X-ray-detected dwarf galaxies in the $L_\mathrm{X}-M_\ast$ plane. 
    X-ray AGNs (red filled circles) are identified following the classification criteria of \citet{Xue2016ApJS..224...15X} and \citet{Luo2017ApJS..228....2L}, and the remaining sources are classified as X-ray galaxies (blue diamonds).
    \textit{Right:} The $M_\ast-z$ distribution for field galaxies (gray points), X-ray AGNs, and X-ray galaxies within the overlapping regions of JWST and the Chandra Deep Fields. The stellar-mass threshold for dwarf galaxies ($M_\ast = 3 \times 10^{9}~M_\odot$) is indicated by the black solid line. Black dashed lines represent the stellar-mass completeness limit for each field, while the orange dashed line indicates the completeness limit for HST. Only mass-complete sources above the black dashed lines are used to derive the star-forming main sequence and for subsequent analyses. 
    The apparent vertical clustering reflects photo-$z$ systematics rather than intrinsic physical overdensities.
    }
    \label{fig:prop1}
\end{figure*}
In our initial selection pass, we isolate candidate dwarf sources with available mass measurements satisfying $M_\ast < 10^{10}\ M_\odot$.
We then employ \texttt{CIGALE version 2025.0} \citep{Boquien2019A&A...622A.103B}, incorporating its X-ray module \citep{Yang2022ApJ...927..192Y}, to perform multi-component SED fitting. 
This step yields robust $M_\ast$ and SFR determinations by leveraging a typical coverage of 18 high-quality (S/N$>5$) photometric bands per source. 
The fitting parameters and adopted grids are summarized in Table~\ref{tab:cigale}. 
Finally, adopting these \texttt{CIGALE}-derived $M_\ast$ as our fiducial masses, we apply a strict dwarf-galaxy mass threshold of $M_\ast < 3 \times 10^9\ M_\odot$, yielding an initial sample of 72 sources. We discuss the potential impact of the AGN contribution to the host-galaxy $M_\ast$ in the final part of Section~\ref{subsec:agn_classification}.

To ensure that our X-ray-detected sample is not biased by mass incompleteness, and to allow for robust comparisons between active and inactive galaxies under identical selection functions, we establish a survey mass-completeness limit. 
We evaluate the stellar mass completeness of our sample using the empirical method developed by \citet{Pozzetti2010A&A...523A..13P}, relying directly on the \texttt{EAZY}-derived stellar masses without applying additional $Q_z$ cuts. 
We adopt a limiting magnitude of $m_{\mathrm{F200W,lim}} = 28.5$, which is approximately $0.5~\mathrm{mag}$ brighter than the typical $5\sigma$ depth measured within a 0\farcs5 diameter aperture across our fields \citep{Valentino2023ApJ...947...20V, Hainline2024ApJ...964...71H}. 
This conservative limit is chosen to account for exposure-time variations across the different observing programs. 
For each detected galaxy, we compute a theoretical limiting stellar mass ($M_{\mathrm{lim}}$), defined as the mass the galaxy would possess if its flux were scaled down to the survey detection limit while preserving its intrinsic mass-to-light ratio. This is calculated as $\log M_{\mathrm{lim}} = \log M_\ast - 0.4(m_{\mathrm{lim}} - m_{\mathrm{obs}})$. 
To avoid biases introduced by bright galaxies, which often exhibit systematically different mass-to-light ratios compared to the faint population, we restrict this calculation exclusively to the faintest 20\% of galaxies within each redshift bin. 
The 90\% mass-completeness threshold for a given redshift interval is then defined as the 90th percentile of this localized $M_{\mathrm{lim}}$ distribution. Adopting this threshold ensures that 90\% of the limiting mass values fall below our cut, guaranteeing a highly host-galaxy mass-complete sample for subsequent analyses.
The resulting mass-completeness curve lies well below the parameter space occupied by most of the X-ray-detected sources, confirming that our sample is not limited by detection sensitivities regarding host galaxy properties. 
Applying this mass-completeness criterion removes only one source from our final sample. As illustrated in Figure~\ref{fig:prop1}, adopting a similar mass-completeness limit based on HST depths would exclude several of the lower-mass and higher-redshift dwarf candidates, precluding the statistically reliable constraints that JWST provides.

Furthermore, to guarantee the reliability of the derived physical parameters, we exclude six sources with poor SED fits ($\chi^2_\nu > 10$) and low photometric-redshift quality ($Q_z>1$), as such fits can severely compromise the derived properties \citep[e.g.,][]{Wasleske2025ApJ...995..160W}. 
As a final verification, we perform source-by-source visual validation of the image cutouts to identify potential cross-matching mismatches, particularly for heavily obscured objects. 
We find several instances of potential mismatches where a close companion resides within the Chandra $1\sigma$ positional uncertainty, but the companion source is optically dark and only detected at $>2~\mu $m. 
For these cases, we rerun \texttt{CIGALE} under the assumption that the X-ray emission originates from the obscured source; this routinely results in a revised mass estimate that exceeds our dwarf mass threshold. 
Consequently, we remove these nine sources from the sample, yielding a final sample of 56 secure X-ray dwarf sources. 

\begin{figure*}[ht]
    \centering
    \includegraphics[width=\linewidth]{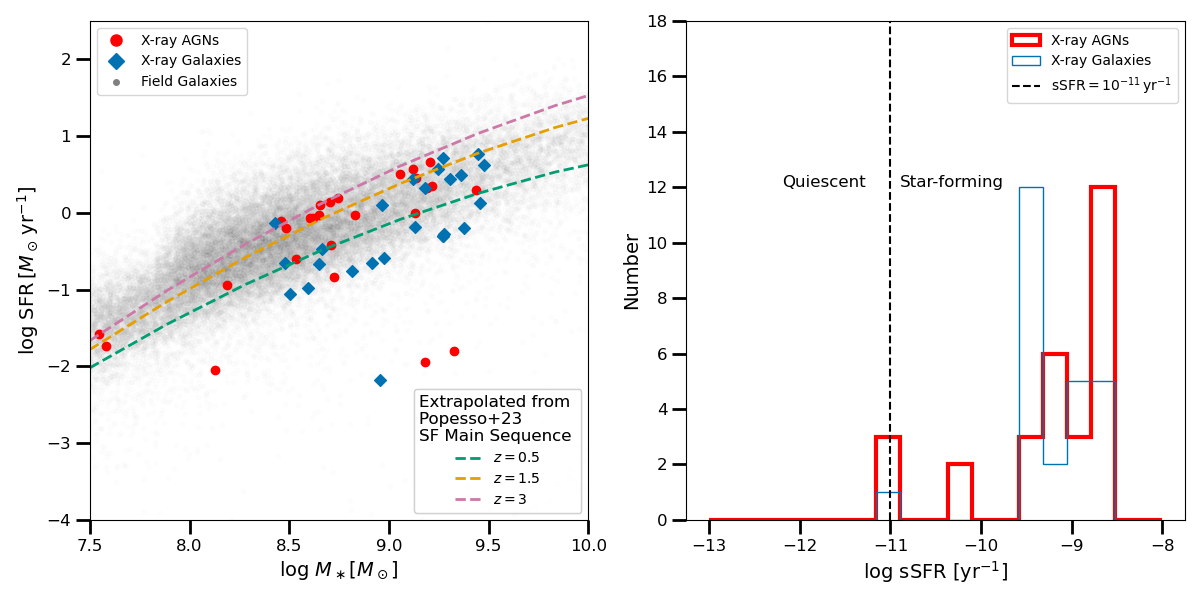}
    \caption{
    \textit{Left:} Location of the X-ray-detected dwarf galaxies in the $\mathrm{SFR}-M_\ast$ plane. Red filled circles denote X-ray AGNs and blue diamonds represent X-ray galaxies, respectively, and gray points show the field dwarf-galaxy population. The star-forming main-sequence relations at $z=0.5$ (teal), $z=1.5$ (orange), and $z=3$ (pink) are shown as dashed lines, extrapolated from \citet{Popesso2023MNRAS.519.1526P}.
    \textit{Right:} Distribution of sSFR for X-ray dwarf galaxies, derived from their best-fit SEDs. Red and blue histograms represent X-ray AGNs and X-ray galaxies, respectively. The majority of the sample is non-quiescent, as defined by a threshold of $\mathrm{sSFR} = 10^{-11}~\mathrm{yr}^{-1}$.} 
    \label{fig:sfr_mass}
\end{figure*}

The physical properties of the full sample are presented in Figure~\ref{fig:prop1}, Figure~\ref{fig:sfr_mass} and Table~\ref{tab:catalog}. 
The left panel of Figure~\ref{fig:prop1} illustrates the $L_\mathrm{X}$--$M_\ast$ plane, demonstrating that the stellar masses and X-ray luminosities span approximately 2--3 orders of magnitude. 
The right panel of Figure~\ref{fig:prop1} displays $M_\ast$ as a function of redshift, highlighting our selected sample relative to the broader field galaxy population and mass-completeness limits. The apparent vertical clustering of field galaxies in photometric redshift is likely dominated by photo-$z$ systematics, such as template degeneracies, limited filter coverage, and redshift-grid effects, rather than intrinsic physical overdensities. 

The left panel of Figure~\ref{fig:sfr_mass} shows the SFR--$M_\ast$ plane, showing that the X-ray dwarf galaxies (red circles and blue diamonds) generally occupy parameter space similar to that of field dwarf galaxies (gray points). 
The majority of them are consistent with star-forming main sequence relations extrapolated from \citet{Popesso2023MNRAS.519.1526P} at similar redshifts, indicating their host galaxies are characterized by ongoing star formation activity. 
The apparent aggregation of X-ray dwarf AGNs (defined in next section) at moderately high SFR and $\log M_\ast\sim 8.6$ is mostly composed of $z>1$ dwarf AGNs. The right panel of Figure~\ref{fig:sfr_mass} shows the distribution of specific SFR (sSFR) for the X-ray dwarf galaxies, where red and blue histograms represent X-ray AGNs and X-ray galaxies, respectively, highlighting their non-quiescent nature given that most of the sources have sSFR$>10^{-11}\,\text{yr}^{-1}$.

\subsection{AGN Classification} \label{subsec:agn_classification}
Most AGN classification methods are optimized for massive galaxies, and their efficacy in identifying dwarf AGNs remains poorly constrained due to our incomplete understanding of dwarf galaxy SEDs and their intrinsic AGN properties \citep{Hainline2016ApJ...832..119H,Zou2023ApJ...950..136Z,Lyu2024ApJ...966..229L,Wasleske2024ApJ...971...68W}.
In low-mass systems, host-galaxy stellar emission frequently dominates the optical-to-mid-infrared SED, heavily diluting the AGN signature. Indeed, as demonstrated by \citetalias{Zou2023ApJ...950..136Z}, a substantial fraction of dwarf AGNs lack a prominent AGN component in their optical-to-infrared SEDs. Requiring a clearly identifiable AGN contribution would therefore preferentially select systems with strong nuclear emission relative to their hosts, potentially excluding genuine AGNs whose emission is diluted by starlight and limiting the representativeness of the dwarf AGN host sample. Consequently, we rely on the broader X-ray AGN classification.

We refine the classification of our sources following the AGN criteria outlined in Section 4.4 of \citet{Xue2011ApJS..195...10X} \citep[also adopted by][]{Xue2016ApJS..224...15X, Luo2017ApJS..228....2L}: (1) an intrinsic X-ray luminosity of $L_\mathrm{X,int}\gtrsim3\times10^{42}~\mathrm{erg~s^{-1}}$ (luminous X-ray sources); (2) an effective power-law photon index of $\Gamma_\mathrm{eff}\leq1.0$ (hard X-ray sources); (3) an X-ray-to-optical flux ratio of $\log(f_\mathrm{X}/f_\mathrm{F606W})>-1$, where $f_\mathrm{X}$ is the detected flux in the full, soft, or hard band (depending on availability) and $f_\mathrm{F606W}$ is the HST/F606W optical flux; and (4) an X-ray-to-NIR flux ratio of $\log(f_\mathrm{X}/f_\mathrm{F200W})>-1.2$, where $f_\mathrm{F200W}$ is the JWST/F200W NIR flux.

Before applying these criteria to the dwarf-galaxy candidates, we verify that they recover the AGNs previously classified by \citet{Xue2016ApJS..224...15X} and \citet{Luo2017ApJS..228....2L}. The four adopted criteria recover approximately 98\% and 95\% of the previously identified AGNs in CDF-N and CDF-S, respectively. Most of the remaining sources were classified in the original catalogs using their excess X-ray emission relative to the 1.4~GHz radio luminosity.

We do not adopt the X-ray-to-radio luminosity-ratio criterion in the present analysis. 
Radio coverage is not uniform between the two fields, and radio emission in low-mass galaxies can include a substantial contribution from star formation, complicating its interpretation as an AGN diagnostic. 
We nevertheless assess whether omitting this criterion affects our final classifications by cross-matching the dwarf-galaxy candidates with the existing radio catalogs of \citet{Miller2013ApJS..205...13M} and \citet{Owen2018ApJS..235...34O}.
For the CDF-S sample, we additionally perform a probabilistic cross-match with the \texttt{MIGHTEE} catalog using \texttt{NWAY}.
This cross-match identifies radio counterparts to two dwarf galaxies, S-316 and S-857, both of which remain classified as X-ray galaxies under our adopted criteria. 
A comparison with the GOODS-S multiwavelength AGN sample of \citet{Lyu2022ApJ...941..191L} similarly indicates that the relevant classification differences are largely associated with this criterion but do not affect our final classifications. 
We therefore conclude that omitting the X-ray-to-radio luminosity-ratio criterion has a negligible effect on the resulting dwarf-AGN sample.

Furthermore, we follow standard practice in the literature by requiring an X-ray excess as a fifth criterion to ensure a reliable sample of dwarf AGNs.
Following \citet{Birchall2020MNRAS.492.2268B}, an X-ray excess AGN is defined as
\begin{align}
    \text{X-ray Excess AGN} : \frac{L_\mathrm{X}}{L_\mathrm{Hot~Gas} + L_\mathrm{XRB}}>3
\end{align}
where $L_{\rm X}$ is the observed 2--10 keV X-ray luminosity, $L_{\rm Hot~Gas}$ is the expected X-ray emission from hot gas, and $L_{\rm XRB}$ is the expected integrated contribution from both low-mass and high-mass X-ray binaries (XRBs). 
This distinction is necessary because both hot gas and XRBs can contribute X-ray emission at levels that may be confused with an AGN.
As shown in the right panel of Figure~\ref{fig:sfr_mass}, the majority of our sources exhibit sSFR $>10^{-11}\ {\rm yr}^{-1}$, which is a common threshold used to distinguish between star-forming and quenched galaxies \citep{Brinchmann2004MNRAS.351.1151B,Fontanot2009MNRAS.397.1776F}. 
Therefore, we estimate the hot-gas emission using the relation from \citet{Mineo2012MNRAS.426.1870M}, which is appropriate for star-formation-dominated galaxies:
\begin{align}
     L^{\text{Hot Gas}}_\mathrm{\text{0.5-2~keV}}=8.3\times 10^{38}~\times \frac{\text{SFR}}{0.63}
\end{align}
where the factor of 0.63 converts the relation from a Salpeter initial mass function (IMF) to the Chabrier IMF utilized in our SED fitting \citep[][see implementation in \citetalias{Zou2023ApJ...950..136Z}]{Speagle2014ApJS..214...15S,Madau2014ARA&A..52..415M}.
For the XRB contribution, we apply the scaling relations from \citet{Lehmer2016ApJ...825....7L}, which account for the redshift evolution and the dependencies of low-mass XRBs (LMXBs) and high-mass XRBs (HMXBs) on $M_\ast$ and SFR, respectively:
\begin{align}
    L^{\text{XRB}}_\mathrm{\text{2-10~keV}} &= L^{\text{LMXB}}_\mathrm{\text{2-10~keV}} + L^{\text{HMXB}}_\mathrm{\text{2-10~keV}} \nonumber \\
    &= 10^{29.37} \times (1+z)^{2.03} \times \frac{M_\ast}{0.94} + \nonumber \\
    &\quad 10^{39.40} \times (1+z)^{1.02} \times \frac{\mathrm{SFR}}{0.94}
\end{align}
A factor of 0.94 is applied to convert the relations from a Kroupa IMF to our adopted Chabrier IMF. We assume a power-law spectrum with a photon index of $\Gamma = 1.8$ for the integrated XRB emission. We confirm that these analytical XRB estimates are consistent with the outputs from the \texttt{CIGALE} XRB module \citep{Yang2020MNRAS.491..740Y}. 
Because dwarf galaxies are typically metal-poor, enhanced contributions from HMXBs can elevate their expected X-ray fluxes \citep{Lehmer2016ApJ...825....7L,Lehmer2021ApJ...907...17L}.
Consequently, sources lying near the X-ray excess threshold warrant caution. 
Nevertheless, the observed excesses are empirically robust: even prior to applying an explicit excess cut, 95\% of all X-ray-detected dwarf candidates exhibit an excess factor of $>2$. 
We adopt a fiducial threshold of $>3$ to identify potential X-ray excess AGNs while maintaining consistency with the literature. 
Varying this boundary to $>5$ alters the final AGN sample by at most five sources. 
Because this minor variation does not qualitatively affect the environmental, morphological, or accretion-distribution results, our conclusions remain robust against plausible XRB scaling uncertainties. 
We further discuss the extreme dwarf galaxy (EDG) AGN candidates in Appendix~\ref{app:edg-agn}.

\begin{figure*}[ht]
    \centering
    \includegraphics[width=\linewidth]{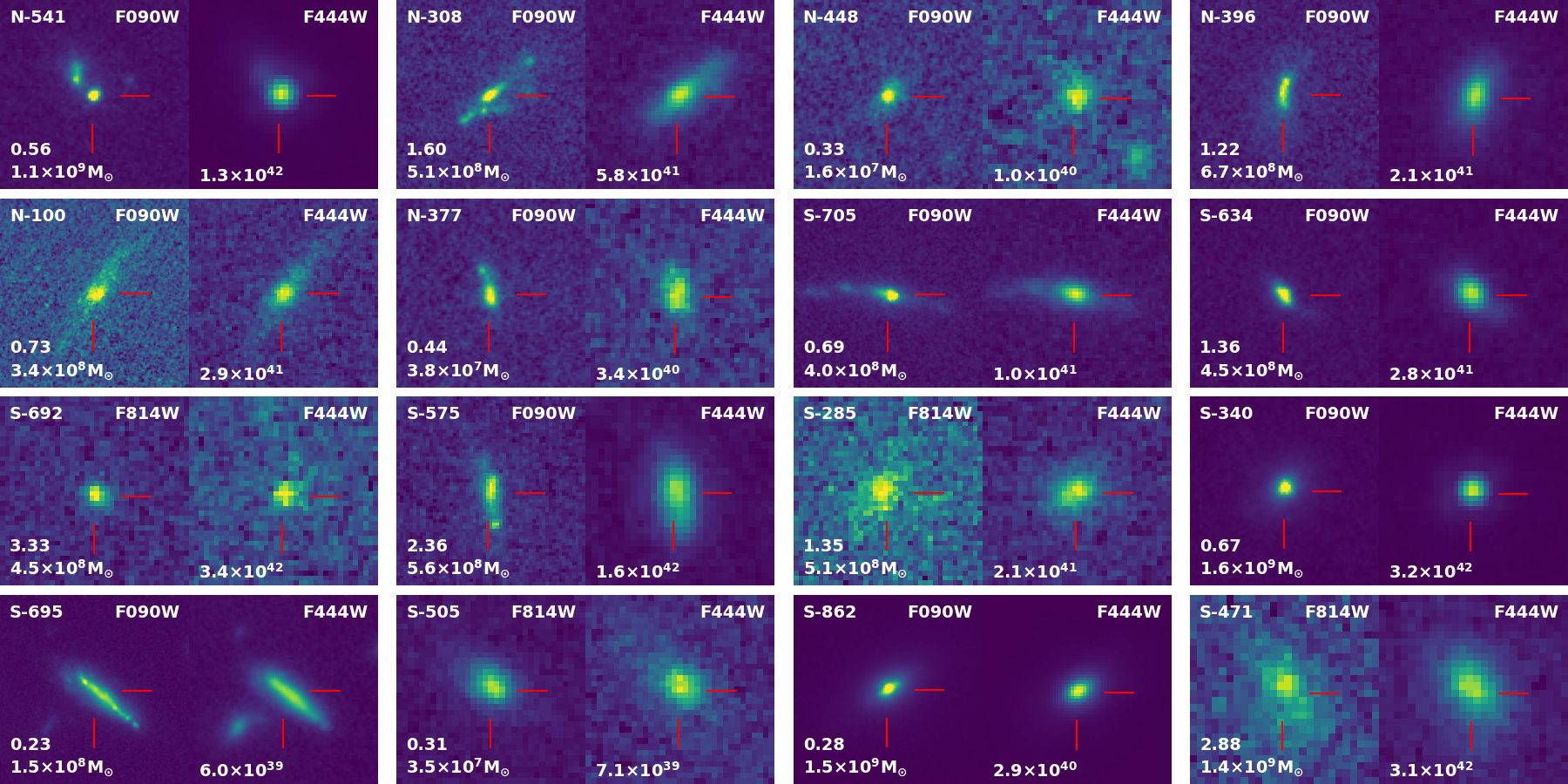}
    \caption{
    Postage-stamp images of 16 X-ray dwarf AGNs identified in this work, arranged in a $4\times4$ grid. Each panel displays a pair of stamps for a single object, labeled with an ``N" or ``S" (denoting the field selection) followed by its ID. The top-right corner of each stamp indicates the corresponding 1 $\mu$m and 4 $\mu$m bands (typically JWST/NIRCam F090W and F444W, respectively). In cases where F090W is unavailable, the nearest available filter is substituted to represent the 1 $\mu$m band. The text in the lower-left corner of the first band provides the $z$ from \texttt{EAZY} (top) and the $M_{\ast}$ from \texttt{CIGALE} SED fitting (bottom); the lower-left corner of the second band displays the $L_\mathrm{X}$ in erg/s. Red crosshairs mark the source positions in the redder JWST NIRCam band.
    } 
    \label{fig:grid_image}
\end{figure*}

\begin{figure*}[ht]
    \centering
    \includegraphics[width=\linewidth]{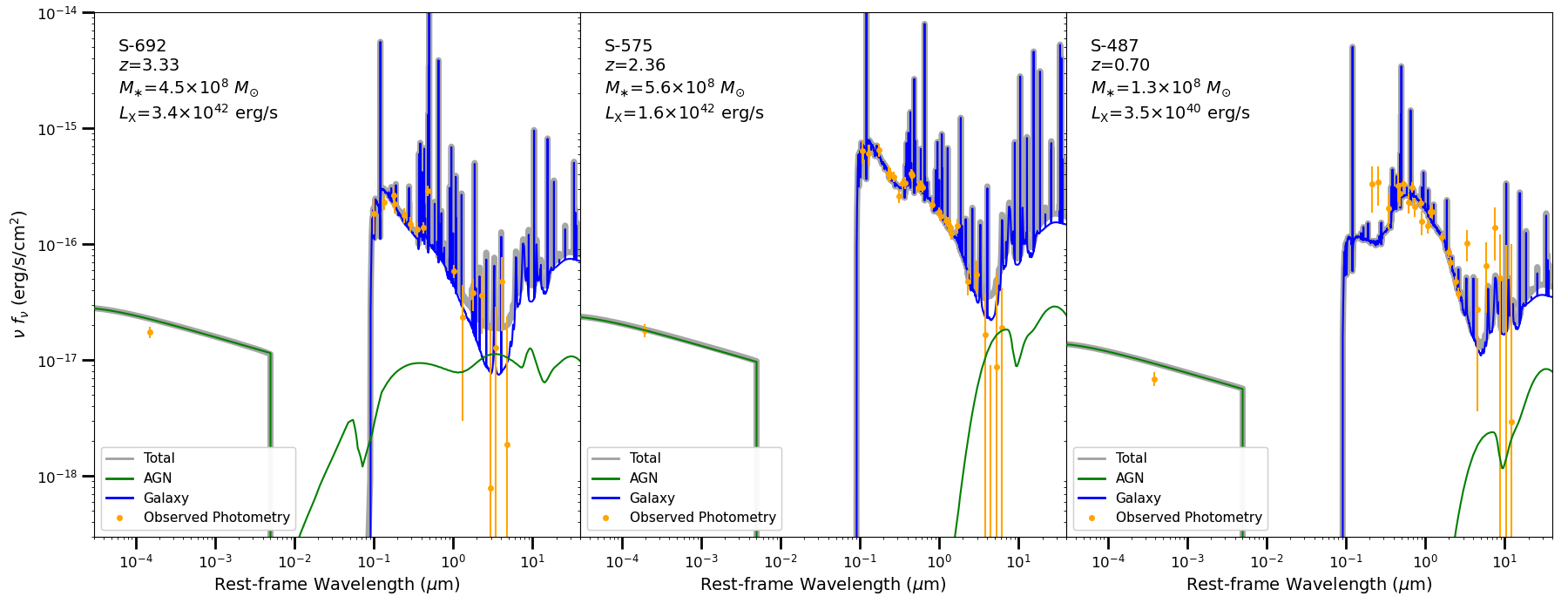}
    \caption{
    Best-fit spectral energy distributions (SEDs) for three representative X-ray dwarf AGNs in our sample. 
    The observed multiwavelength photometry is represented by orange circles with $1\sigma$ error bars, with downward-pointing triangles indicating upper limits. The thick gray solid line represents the total best-fit model SED. This total model is further decomposed into the emission from the host galaxy (blue solid line) and the AGN component (green solid line). Key physical properties of the sources, including the redshift ($z$), stellar mass ($M_\ast$), and X-ray luminosity ($L_\mathrm{X}$), are annotated in the upper-left corner of each panel.
    } 
    \label{fig:sed_plot}
\end{figure*}

Sources satisfying any of the five aforementioned criteria are classified as AGNs, yielding a final sample of 31 X-ray-detected dwarf AGNs and 25 X-ray-detected dwarf galaxies. 
We caution that the latter should not be interpreted as a purely non-AGN population. 
Rather, this subsample comprises X-ray-detected dwarf galaxies whose observed X-ray luminosities and flux ratios do not satisfy our conservative AGN criteria after accounting for the expected XRB and hot-gas contributions. 
It is likely that some of these sources host low-luminosity, variable, or heavily obscured AGNs that fall below our classification threshold \citep[e.g.,][]{Zheng2017ApJ...849..127Z}.
Figure~\ref{fig:grid_image} presents a subset of the X-ray dwarf AGNs defined in this work, displaying paired short-wavelength ($\sim1~\mu$m) and long-wavelength ($\sim4~\mu$m) postage-stamp images. 
The central target positions are marked with red crosshairs. Each pair is annotated with the source ID in the upper left (where ``N" and ``S" denote the GOODS-N and GOODS-S fields, respectively); the redshift $z$ and stellar mass $M_\ast$ in the lower left of the first panel; and the X-ray luminosity $L_\mathrm{X}$ in the lower left of the second panel. 
Furthermore, Figure~\ref{fig:sed_plot} presents the rest-frame multiwavelength spectral energy distributions (SEDs) for three example sources. 
These plots demonstrate our SED decomposition, showing that the combination of host-galaxy stellar emission (blue curves) and AGN emission (green curves) yields a best-fit total model (gray curves) in good overall agreement with the observed photometry (orange circles). 

The X-ray-selected dwarf AGNs in our sample generally exhibit weak optical and near-infrared (NIR) AGN signatures. Despite their X-ray luminosities, most sources show neither prominent nuclear point sources in the imaging data nor substantial AGN contributions in their SEDs. 
For example, Figures~\ref{fig:grid_image} and \ref{fig:sed_plot} include several of the most X-ray-luminous systems in our sample ($L_\mathrm{X}\sim3\times10^{42}~\mathrm{erg\,s^{-1}}$), yet their optical/NIR morphologies remain largely host-dominated and their SEDs show only modest evidence for AGN emission. We discuss the details later in Section~\ref{sec:nuclear_ps}.


To ensure that our core demographic and evolutionary findings are not artificially driven by the strict dwarf-galaxy stellar-mass threshold of $3 \times 10^9~M_\odot$, we systematically examine the sensitivity of our results to the expected scatter in stellar-mass estimates. 
As previously noted, the presence of an AGN can affect the $M_\ast$ estimation, potentially yielding unreasonably low $M_\ast$ values for systems where the AGN dominates the optical-to-NIR SED. 
We adopt a fiducial uncertainty of $\sim0.3$ dex for SED-based stellar masses. 
To verify this, we additionally run \texttt{CIGALE} without the AGN module and compare the resulting $M_\ast$ estimates to those derived with the AGN component included. 
For the vast majority of our dwarf candidates, the mass difference is well within $\sim0.3$ dex. 
Furthermore, previous studies of low-mass galaxies have demonstrated that parametric SFH assumptions can introduce systematic offsets of up to $\sim0.4$ dex if the true SFHs are bursty, so our stellar masses may be subject to additional SFH-dependent systematics \citep[e.g.,][]{Kurczynski2016ApJ...820L...1K,Lower2020ApJ...904...33L,DeLosReyes2025ApJ...989...91D}. 
To account for these combined uncertainties, we test the robustness of our sample selection by shifting the mass threshold by $\pm 0.3$ dex. 
Expanding the threshold by $+0.3$ dex increases the total sample to 80 sources (44 AGNs) , while a more conservative $-0.3$ dex threshold reduces the sample to 38 sources (25 AGNs). 
We repeat the analyses described below on these boundary samples and verify that our principal results remain statistically unchanged regardless of the adopted mass threshold.

\section{Method}
\subsection{Morphological Information: \textit{GMCA}}\label{sec:gmca}
To quantitatively characterize the structural properties of the galaxies in our sample, we measure their nonparametric morphological parameters. 
Unlike parametric fitting (e.g., single- or multi-component S\'ersic profiles), nonparametric morphologies do not rely on predefined analytical assumptions regarding the light distribution. 
This makes them highly robust and advantageous for characterizing irregular, asymmetric, or merging galaxies that exhibit complex, disturbed structures.

For this analysis, we utilize the open-source Python package \texttt{statmorph}\footnote{\url{https://statmorph.readthedocs.io}} \citep{Rodriguez2019MNRAS.483.4140R}. 
For each galaxy of interest, we prepare cutouts of background-subtracted science images, associated sigma images for error propagation and S/N estimation, a segmentation map, and point spread function (PSF) images in each filter band.   
\texttt{statmorph} extracts a suite of standard morphological diagnostics, primarily encompassing the CAS system \citep{Conselice2003ApJS..147....1C} and the Gini-$M_{20}$ statistics \citep{Lotz2004AJ....128..163L}  computed within 1.5 $r_\mathrm{petro}$ \citep{Conselice2003ApJS..147....1C,Lotz2004AJ....128..163L}. We focus on the \textit{GMCA} parameters:

\textit{Gini Coefficient ($G$):} 
$G$ quantifies the inequality of the flux distribution among the pixels assigned to the galaxy \citep{Lotz2004AJ....128..163L}.
A value of $G = 0$ indicates a perfectly uniform surface-brightness distribution, whereas $G = 1$ denotes that all flux is concentrated in a single pixel.
    
\textit{$M_{20}$ statistic:} 
$M_{20}$ represents the normalized second-order moment of the brightest 20\% of the galaxy's pixels \citep{Lotz2004AJ....128..163L}. Unlike the central concentration index, $M_{20}$ is highly sensitive to the spatial distribution of the brightest regions and can effectively identify multiple bright nuclei or off-center starbursts.

\textit{Concentration ($C$):} $C$ traces the prominence of the central bulge relative to the extended disk, defined as $C = 5 \log_{10}(r_{80}/r_{20})$, where $r_{80}$ and $r_{20}$ are the circular apertures containing 80\% and 20\% of the total galaxy flux, respectively. 
    
\textit{Asymmetry ($A$):} 
Computed by subtracting a $180^\circ$-rotated image of the galaxy from its original image, normalized by the original total flux, and subtracting background asymmetry.
High $A$ values typically indicate spatial disturbances driven by ongoing mergers, tidal interactions, or highly asymmetric star-formation activities.

To ensure the reliability of the derived morphological metrics, we restrict our analysis to sources satisfying the following selection criteria: (i) A \texttt{statmorph} quality flag of $\texttt{flag} \leq 1$. (ii) Measurements that are robust against artifacts or edge-effects in the segmentation mask generation. (iii) A 20\% light radius ($r_{20}$) strictly greater than half of the point spread function's full width at half maximum ($r_{20} > 0.5 \times \mathrm{FWHM_{PSF}}$). (iv) An average signal-to-noise ratio per pixel (\texttt{sn\_per\_pixel}) greater than 2.5.

We apply these quality checks to guarantee that the resulting morphological parameters are robust enough to provide meaningful statistical constraints. It is important to note that not all statistics are successfully derived across every filter. For this analysis, we extract parameters using all available broadband imaging spanning 0.4--5 $\mu$m.

\subsection{AGN Accretion Distribution Modeling, AGN Fraction, and $\overline{\mathrm{BHAR}}$}\label{sec:accretion_modeling}
We model the X-ray specific accretion-rate distribution of dwarf galaxies to infer the observable active fraction ($f_\mathrm{AGN}$) over the luminosity range probed by our sample. 
This framework describes the probability that a galaxy with a given stellar mass and redshift hosts an X-ray AGN of a given luminosity, while accounting for survey sensitivity. 
The specific accretion-rate distribution is defined as the conditional probability density per unit $\log L_\mathrm{X}$ of a galaxy at a given $(M_\ast,z)$ hosting an AGN with an intrinsic  2--10 keV $L_\mathrm{X}$ \citep{Aird2012ApJ...746...90A,Aird2018MNRAS.474.1225A,Yang2018MNRAS.475.1887Y,Zou2023ApJ...950..136Z}. 
Following established literature, we assume a power-law functional form:
\begin{align}
    p\left(L_\mathrm{X}\,|\, M_\ast,z \right) = A~\left(\frac{M_\ast}{M_0}\right)^{\gamma_M}~\left(\frac{L_\mathrm{X}}{L_0}\right)^{\gamma_L}~\left(\frac{1+z}{1+z_\mathrm{0}}\right)^{\gamma_z}\label{eq:pLx}
\end{align}
where {$A,\gamma_M,\gamma_L,\gamma_z$} are the free parameters governing normalization and slope of the accretion distribution. The scaling factors are fixed at $(M_0,L_0,z_0)=(10^{11}~M_\odot,10^{43}~\mathrm{erg~s^{-1}},0.6)$, following \citet{Aird2012ApJ...746...90A}. While this parameterization was initially developed for massive galaxies, recent studies have demonstrated its validity and efficacy when extended to the dwarf-galaxy regime \citep{Birchall2020MNRAS.492.2268B, Birchall2022MNRAS.510.4556B, Zou2023ApJ...950..136Z}. 

Prior to parameter fitting, we must ensure our input sample is unbiased by mass incompleteness (as addressed in Section~\ref{sec:sed_fitting_dwarf_selection}) and survey depth limits. 
To correct for X-ray survey incompleteness, we adopt the methodology detailed in Section~3.1.1 of \citet{Zou2024ApJ...964..183Z}. 
Specifically, we model the probability, $P_\mathrm{det}(f_\mathrm{X})$, that a source with a 2--10 keV flux ($f_\mathrm{X}$) is detected by the X-ray survey using the following functional form:
\begin{align}
    P_\mathrm{det}(f_\mathrm{X}) = \frac{1}{2}\left[\mathrm{erf}(b(\log f_\mathrm{X}-a))+1 \right]\label{eq:pdet_fx}
\end{align}
where $a$ and $b$ are parameters governing the shape of the detection probability curve. As established by \citet{Zou2024ApJ...964..183Z}, these $(a, b)$ values were derived by convolving $P_\mathrm{det}(f_\mathrm{X})$ with the well-calibrated 2--10 keV $\log N - \log S$ relation from \citet{Georgakakis2008MNRAS.388.1205G}, and fitting the result to match the observed $f_\mathrm{X}$ distribution. The best-fit (a,b) for our regions of interest are (-16.32,3.58) and (-16.74, 7.53) for GOODS-N and GOODS-S, respectively.

Next, we construct the log-likelihood function for the probability distribution $p\left(L_\mathrm{X}| M_\ast,z \right)$ as follows:
\begin{align}
    \ln \mathcal{L}_{k} = -N_\mathrm{mdl} + \sum^{n_\mathrm{AGN}}_{i=1} \ln p\left(L_{\mathrm{X},i}\,| M_{\ast,i},z_i \right)
\end{align}
where the summation index $i$ runs over X-ray AGNs. The term $N_\mathrm{mdl}$ is the expected number of X-ray-detected active dwarf galaxies given the survey depth, and index $k$ denotes the field (i.e., CDF-N or CDF-S). The total log-likelihood is the sum of the individual field log-likelihoods. 

$N_\mathrm{mdl}$ is calculated by convolving the detection probability, $P_\mathrm{det}(f_\mathrm{X})$, with $p(L_\mathrm{X}\,|\, M_\ast,z)$. 
Limiting our estimation to the physically relevant luminosity range, we define this as the following:
\begin{align}
N_\mathrm{mdl} = \sum^{n_\mathrm{gal}}_{j=1} \int^{\log L_\mathrm{X,max}}_{\log L_\mathrm{X,min}} p\left(L_\mathrm{X} \,|\, M_{\ast,j},z_j \right) P_\mathrm{det}(f_\mathrm{X}) ~ d\log L_\mathrm{X}, \label{eq:n_mdl}
\end{align}
where $f_\mathrm{X}=f_\mathrm{X}(L_\mathrm{X},z)$ is the intrinsic X-ray flux for a source with X-ray luminosity $L_\mathrm{X}$ at redshift $z$, assuming a power-law spectrum with a photon index of $\Gamma=1.8$. 
The bounds $\log L_\mathrm{X,min}$ and $\log L_\mathrm{X,max}$ correspond to the $L_\mathrm{X}$ range probed in this study. 
By defining the characteristic luminosity anchor point 
$\mu_j = \log \left(L_{\mathrm{X},j}\right) = a + \log(4\pi d^2_{\mathrm{L},j})$ 
and applying the variable transformation $x = b\left(\log L_\mathrm{X}-\mu_j\right)$, the integrand can be mapped to a standard analytical form. 
Substituting $c = \frac{\gamma_L \ln(10)}{b}$, we evaluate the indefinite integral using the identity from \citet{Ng1969table},
\begin{align}
\int e^{cx} [\text{erf}(x) + 1]  dx = \frac{1}{c} \left\{ e^{cx} [\text{erf}(x) + 1] - e^{\frac{c^2}{4}} \text{erf}\left(x - \frac{c}{2}\right) \right\}.
\end{align}
To ensure numerical stability when evaluating the definite boundaries, we utilize the identities $\text{erf}(x) = 1-\text{erfc}(x)$ and $\text{erfcx}(x) = e^{x^2}\text{erfc}(x)$. 
Because the constant terms cancel over the definite integral limits, the negative boundary term transforms into $+e^{c^2/4}\,\text{erfc}\left(x-\frac{c}{2}\right) = e^{cx-x^2}\text{erfcx}\left(x-\frac{c}{2}\right)$. 
Defining the numerically stable bounding function $\mathcal{H}(x) = e^{cx} [\text{erf}(x) + 1] + e^{cx-x^2}\text{erfcx}\left(x - \frac{c}{2}\right)$, we can reduce Equation~\ref{eq:n_mdl} to the closed-form analytical solution:
\begin{align}
N_\mathrm{mdl} = \sum^{n_\mathrm{gal}}_{j=1} \frac{p\left(L_{{\mathrm{X},j}}\,|\, M_{\ast,j},z_j \right)}{2\gamma_L \ln 10} \left[ \mathcal{H}\left(x_{\mathrm{max},j}\right) - \mathcal{H}\left(x_{\mathrm{min},j}\right) \right]. \label{eq:n_mdl_analytic}
\end{align}

We adopt flat priors on the free parameters $(A,\gamma_M,\gamma_L,\gamma_z)$ and utilize the \texttt{emcee} \citep{Foreman-Mackey2013PASP..125..306F} package to perform Markov Chain Monte Carlo (MCMC) sampling of the posterior distribution. 
To robustly propagate observational uncertainties into our parameter estimation, we dynamically draw $z$ and $M_\ast$ values from their respective probability density functions at each step of the MCMC sampler. 
Specifically, for sources lacking spectroscopic confirmation, photometric redshifts are sampled from an asymmetric Gaussian distribution defined by the 16th, 50th, and 84th percentiles ($z_{16}$, $z_{50}$, and $z_{84}$) of the photo-$z$ posterior. 
Similarly, to account for the inherent statistical and systematic uncertainties in deriving stellar masses for dwarf galaxies, we sample $M_\ast$ from a symmetric Gaussian distribution centered on the best-fit mass with a conservative standard deviation of $0.3$ dex. 
By introducing this randomization at every likelihood evaluation, we ensure that the asymmetric photo-$z$ probability density functions and the stellar-mass uncertainties are fully marginalized into the final posterior constraints.

The AGN fraction, $f_\mathrm{AGN}$, represents the integrated probability of a galaxy hosting an active BH above a specific luminosity threshold. It is calculated by integrating the $p(L_\mathrm{X} \,|\, M_\ast, z)$ over the chosen X-ray luminosity range,
\begin{align}
f_\mathrm{AGN} &= \int_{L_\mathrm{min}}^{L_\mathrm{max}}p(L_\mathrm{X}\,|\,M_\ast,z) \,d\log L_\mathrm{X}.\label{eq:agn_fraction}
\end{align}

To quantify the sustained growth of SMBHs while minimizing the impact of stochastic, short-term variability, we utilize the long-term average BH accretion rate ($\overline{\mathrm{BHAR}}$). This value represents the mean growth rate for a population of galaxies at a fixed $M_\ast$ and $z$. The accretion rate is fundamentally linked to the bolometric luminosity ($L_\mathrm{bol}$), which we estimate by applying a bolometric correction ($k_\mathrm{bol}=k_\mathrm{bol}(L_\mathrm{X})$, \citealt{Duras2020A&A...636A..73D}) to the observed $L_\mathrm{X}$:
\begin{align}
\mathrm{\overline{BHAR}} &= \int_{L_\mathrm{min}}^{L_\mathrm{max}}\left[\frac{(1-\epsilon) k_\mathrm{bol}L_\mathrm{X}}{\epsilon c^2}\right] \cdot \left[p(L_\mathrm{X}\,|\,M_\ast,z)\right] \,d\log L_\mathrm{X}\label{eq:bhar}
\end{align}
where $\epsilon$ is the radiative efficiency (typically assumed to be $\approx 0.1$). 
Because this conversion depends on the adopted bolometric correction and radiative efficiency, we interpret the derived $\overline{\mathrm{BHAR}}$ values as standardized, model-dependent indicators of population-averaged black-hole growth rather than direct measurements of the physical accretion rates of individual low-mass BHs.

\section{Results and Discussion}
\subsection{X-ray Dwarf AGN Source Properties}
\subsubsection{Number Densities}
The combined GOODS-N and GOODS-S footprint utilized in this work covers an effective area of $\simeq$ 117 arcmin$^2$, comprising $\simeq$ 57 arcmin$^2$ in GOODS-N and $\simeq$ 60 arcmin$^2$ in GOODS-S. 
Within this area, we identify 56 securely X-ray-detected dwarf sources, consisting of 31 X-ray AGNs and 25 X-ray galaxies. 
The corresponding observed surface density of X-ray-selected dwarf AGNs is therefore
$\Sigma_\mathrm{AGN} = 0.26 ~\mathrm{arcmin^{-2}}$ (equivalently, $9.54  \times 10^2~\mathrm{deg^{-2}}$). 

Because the central 7~Ms X-ray observations in GOODS-S are considerably deeper than the 2~Ms observations in GOODS-N, the raw combined surface density is inherently influenced by depth variation. We use Equation~\ref{eq:pdet_fx} to recalculate the expected detection probabilities of the GOODS-S sources by substituting the shallower GOODS-N survey parameters. 
By homogenizing the survey depths in this manner, we find that both fields yield a consistent expected surface density for X-ray-selected dwarf AGNs of $\Sigma_\mathrm{AGN} \sim 0.2 ~\mathrm{arcmin^{-2}}$ ($7 \times 10^2~\mathrm{deg^{-2}}$).
Finally, if the full X-ray-detected dwarf population is considered, including both X-ray AGNs and X-ray galaxies, the total observed surface density increases to $0.48 ~\mathrm{arcmin^{-2}}$ ($1.7 \times 10^3~\mathrm{deg^{-2}}$). 

These values should be interpreted as observed surface number densities for the combination of Chandra sensitivities, JWST survey depths, and dwarf-galaxy selection adopted here, rather than as intrinsic sky surface number densities of all dwarf AGNs. 
The observed surface number density depends strongly on X-ray sensitivity, host-galaxy mass completeness, redshift coverage, and the adopted AGN-classification criteria. 
In particular, the CDFs probe substantially fainter X-ray fluxes than wide-area surveys, and therefore recover a higher surface number density of low-luminosity and high-redshift dwarf AGN candidates.
Direct comparisons with previous dwarf AGN studies are therefore challenging. 
For example, \citet{Mezcua2018MNRAS.478.2576M} identified 40 X-ray-selected AGNs in dwarf galaxies at $z \lesssim$ 2.4 in the 2.2 deg$^2$ Chandra COSMOS-Legacy field, corresponding to a much lower raw surface number density of $\Sigma_\mathrm{AGN} \sim 18~\mathrm{deg}^{-2}$, but over a substantially wider and shallower survey area. 
At lower redshift, \citet{Birchall2020MNRAS.492.2268B} identified 61 X-ray-selected AGNs at $z \sim 0.03$ in the overlap between SDSS and 3\textit{XMM}, corresponding to $\Sigma_\mathrm{AGN} \sim 5.4~\mathrm{ deg}^{-2}$. 
Similarly, \citetalias{Zou2023ApJ...950..136Z} identified 73 X-ray-selected dwarf AGNs at $z < 1$ across three LSST Deep-Drilling Fields covering a total area of 13 deg$^2$, corresponding to $\Sigma_\mathrm{AGN} \sim 5.6~\mathrm{deg}^{-2}$. 
These lower surface densities are expected because the surveys cover much larger areas but are shallower than the CDFs, and because their redshift and stellar-mass selections differ from those adopted here.

Samples selected outside the X-ray band probe different AGN populations and should not be compared directly to our X-ray-selected surface density. 
For example, \citet{Mountrichas2025A&A...702A...6M} compiled 1451 dwarf AGNs selected using emission-line diagnostics and infrared colors at $z =$ 0.5--0.9 from the spectroscopic VIPERS survey, which covers 24 deg$^2$, corresponding to $\Sigma_\mathrm{AGN} \sim 60~\mathrm{deg}^{-2}$. 
Similarly, \citet{Kaviraj2026MNRAS.545f1918K} identified 30 dwarf AGNs in the $\sim$ 1 deg$^2$ VST-COSMOS survey using optical variability, corresponding to $\Sigma_\mathrm{AGN} \sim 30~\mathrm{deg}^{-2}$. 
These optical, infrared, and variability-selected samples are sensitive to different accretion states, obscuration regimes, variability amplitudes, and host-galaxy properties, and therefore are complementary to the X-ray-selected population considered here.

The GOODS-S multiwavelength AGN sample of \citet{Lyu2022ApJ...941..191L} provides a more directly relevant comparison because of its substantial footprint overlap with this work and its broad X-ray-to-radio SED-based AGN selection. 
Within a similar GOODS-S area, their dwarf AGN surface density is $\approx$ 1.6 arcmin$^{-2}$. 
Based on their AGN-selection Venn diagram, the X-ray-detected subset appears to comprise roughly one third of the full multiwavelength AGN population. This fraction should be treated only as an approximate estimate, rather than as a quantitatively matched selection measurement. Applying this approximate fraction to their dwarf AGN surface density implies an X-ray-detected dwarf AGN surface density of $\approx$ 0.5 arcmin$^{-2}$, broadly consistent with the surface density obtained here when both X-ray AGNs and X-ray galaxies are included.

Therefore, the high observed surface density measured in this work does not imply tension with previous dwarf AGN searches. 
Instead, it reflects the unique depth of the Chandra Deep Fields combined with JWST-enabled dwarf-galaxy identification at faint magnitudes and higher redshifts. 
A direct comparison with intrinsic number densities from other surveys requires a forward model that applies matched stellar-mass limits, redshift intervals, X-ray luminosity thresholds, sensitivity functions, and AGN-selection criteria. 
We defer such a full luminosity-function analysis to future work and use the present surface-density estimate primarily as an empirical characterization of the X-ray-selected dwarf AGN yield in the deepest Chandra fields.

\subsubsection{Hardness Ratio}\label{sec:hardness_ratio}
\begin{figure}[h]
    \centering
    \includegraphics[width=\linewidth]{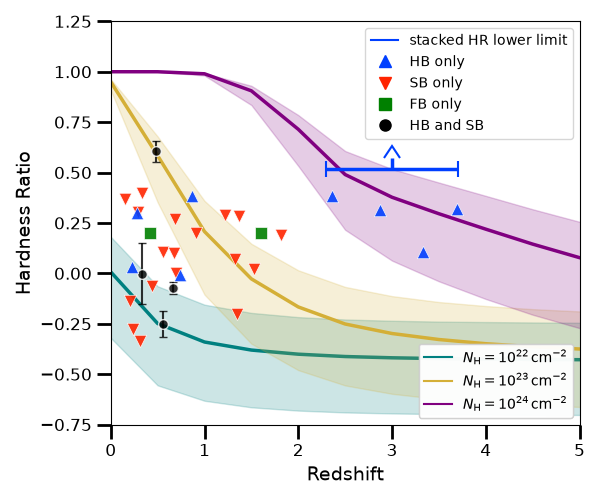}
    \caption{Hardness ratio (HR) as a function of redshift for the active dwarf galaxies in our sample. The legend denotes the detection status of the sources in the soft band (SB), hard band (HB), and full band (FB). Red downward triangles (detected in the SB but not the HB) represent 90\% HR upper limits, and blue upward triangles (detected in the HB but not the SB) denote 90\% HR lower limits. 
    Green squares represent sources detected only in the full band, for which we assign a nominal HR value of 0.2.
    Black circles with error bars indicate the median HR values and their associated $1\sigma$ uncertainties for sources detected in both bands. 
    The shaded regions indicate the expected theoretical relations for redshifted absorbed power laws with photon indices ($\Gamma$) varying between 1.4 and 2.6, evaluated at three intrinsic column densities: $N_\mathrm{H}=10^{22}~\mathrm{cm}^{-2}$ (teal), $10^{23}~\mathrm{cm}^{-2}$ (gold), and $10^{24}~\mathrm{cm}^{-2}$ (purple). The corresponding solid curves denote the expected relations for a fixed photon index of $\Gamma = 1.8$. 
    At $z < 2$, our X-ray-detected AGNs generally lack heavy X-ray obscuration ($N_\mathrm{H} < 10^{23}~\mathrm{cm}^{-2}$). 
    While the stacked HR lower limit of 0.52 at $z > 2$ qualitatively favors the presence of obscuration, our small sample size precludes robust statistical constraints.}
    \label{fig:hardness_z}
\end{figure}
We present the distribution of hardness ratios (HRs) of the dwarf AGNs in Figure~\ref{fig:hardness_z}. We derive the HR using 
\begin{align}
\mathrm{HR} = \frac{C_\mathrm{HB}-C_\mathrm{SB}}{C_\mathrm{HB}+C_\mathrm{SB}}
\end{align}
where $C_\mathrm{HB}$ and $C_\mathrm{SB}$ denote the aperture-corrected, background-subtracted net counts in the hard band (2--7 keV) and soft band (0.5--2 keV), respectively. 
The net counts are extracted directly from the \citet{Xue2016ApJS..224...15X} and \citet{Luo2017ApJS..228....2L} catalogs. When a source is detected in only one band, we adopt the 90\% confidence upper or lower limits provided by the respective catalogs. 
We estimate the HR for three intrinsic column densities using \texttt{PIMMS-v4.15}\footnote{\url{https://heasarc.gsfc.nasa.gov/docs/software/tools/pimms.html}} \citep{Mukai1993Legac...3...21M}. 
To account for the evolution of Chandra's instrumental response, we adopt the Cycle 3 and Cycle 13 ACIS-I effective-area curves as representative of the observing epochs of CDF-N and CDF-S, respectively. We combine the predicted soft- and hard-band counts using the respective survey exposures of 2 and 7 Ms to obtain exposure-weighted HR curves. 
The central curves assume a photon index of $\Gamma=1.8$, while the shaded regions encompass varying ranges of $\Gamma$=1.4--2.6. These tracks provide qualitative guides to the effects of X-ray obscuration rather than precise estimates of individual column densities.

At $z < 2$, most of our X-ray dwarf AGNs are consistent with modest obscuration, exhibiting HR values that broadly correspond to $N_\mathrm{H}<10^{23}~\mathrm{cm^{-2}}$. 
This behavior is qualitatively consistent with previous X-ray studies of dwarf AGNs at $z < 1$ \citep[e.g.,][]{Mezcua2018MNRAS.478.2576M,Zou2023ApJ...950..136Z}, which generally find that the detected X-ray dwarf AGN population is dominated by relatively unobscured sources. 
One possible explanation is that efficiently accreting BHs in dwarf hosts with high $L_\mathrm{X}$ may clear out circumnuclear gas through radiative feedback, leading to lower obscuring columns and a higher fraction of unobscured systems than is typically observed among more massive AGNs \citep[e.g.,][]{Ricci2017Natur.549..488R}.

In contrast, the $z > 2$ sources in our sample are preferentially detected in the hard band, and their HR lower limits are broadly consistent with obscured X-ray spectra. 
Given the small number of high-redshift sources and the limited photon statistics, these HR constraints should not be interpreted as direct measurements of $N_\mathrm{H}$. 
To better constrain the average X-ray properties of these high-redshift sources and to overcome the sensitivity limits of individual detections, we perform a stacking analysis of the four $z > 2$ sources (all lie in the CDF-S). 
We extract $30\arcsec \times 30\arcsec$ image cutouts from the data products of \citet{Luo2017ApJS..228....2L}. 
All four sources lie at small off-axis angles ($\theta < 8\arcmin$), where the results are not strongly affected by Chandra PSF degradation or elevated background noise. 
We stack the image counts and perform aperture photometry measured within circular apertures corresponding to the 90\% encircled-energy fraction of the local Chandra PSF. The local background is estimated from concentric annuli with inner and outer radii of 2.5 and 3.5 times the source aperture radius, respectively.
The stacked hard- and full-band net counts yield marginal detections at the $\sim2.5\sigma$ level, whereas the soft-band stack remains undetected. 
Incorporating the hard-band counts and the 90\% upper limit on the soft-band counts yields a stacked HR constraint of $\mathrm{HR}_\mathrm{stacked} \geq 0.52$.

This lower limit is suggestive of heavy X-ray obscuration and is qualitatively consistent with a scenario in which high-redshift accreting BHs are increasingly obscured by gas on host-galaxy scales.
For example, \citet{Gilli2022A&A...666A..17G} argued that the interstellar-medium column densities of massive galaxies increase rapidly toward earlier cosmic epochs, and these become comparable to X-ray-derived AGN absorbing columns by $z \gtrsim 2$. 
Although their analysis focused on galaxies substantially more massive than the dwarf hosts considered here, the same physical trend of increasing gas fractions and compact galaxy sizes at high redshift may contribute to the harder observed X-ray colors of our $z > 2$ dwarf AGN candidates. 
We therefore interpret the high-redshift HR distribution as tentative evidence that some early dwarf AGNs may experience enhanced host-galaxy or nuclear obscuration. However, deeper X-ray observations and direct spectral modeling will be required to distinguish among intrinsic absorption, redshift-dependent spectral effects, and selection biases.

\subsubsection{Local Host Environments}\label{sec:local_host_environments}
\begin{figure*}[ht]
    \centering
    \includegraphics[width=\linewidth]{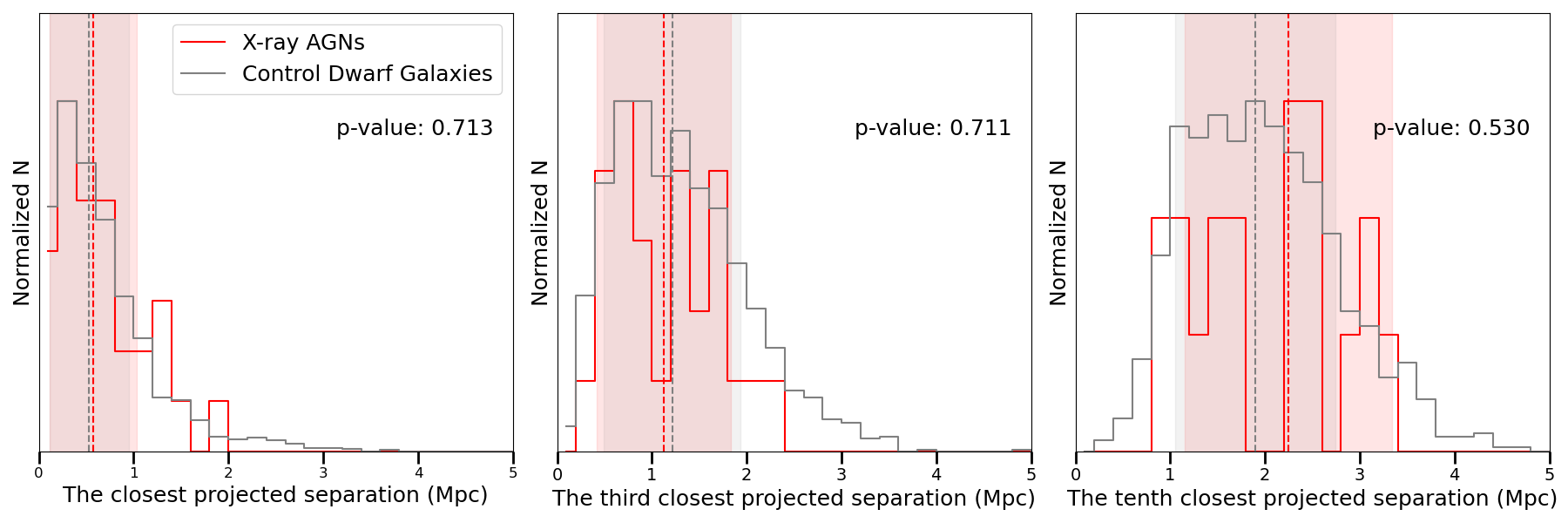}
    \caption{
    \textit{Left, middle, and right panels:} Distributions of the closest, third-closest, and tenth-closest projected separations between X-ray AGNs (red histograms; with stellar mass $M_\ast$ and redshift $z$) and neighboring massive galaxies ($M_\ast^{\mathrm{MG}} > 10^{10.5}~M_\odot$) within a relative redshift slice of $z^{\mathrm{MG}} = z \pm 1.5(1+z)\sigma_{\Delta z / (1+z)}$. 
    These projected separations probe the local environment on scales ranging from $\sim 0.1$ to $5~\mathrm{Mpc}$. Gray histograms represent the baseline separations derived from a control sample of 2000 randomly selected field dwarf galaxies, which are closely matched in stellar mass ($\log M_\ast^{\mathrm{C}} = (1\pm0.1)\log M_\ast$) and redshift ($z^{\mathrm{C}} = z \pm 0.075(1+z)$). Red and gray dashed lines, along with their respective shaded regions, indicate the median values and $1\sigma$ scatter for each distribution. 
    We find no statistically significant evidence that the X-ray AGNs occupy different projected environments from the dwarf-galaxy control sample.
    This is corroborated by two-sample K-S tests, which yield $p > 0.05$ across all panels (as labeled).}
    \label{fig:local_environment}
\end{figure*}

The relationship between AGNs and their host environments within the dwarf-galaxy regime remains poorly understood \citep[e.g.,][]{Greene2020ARA&A..58..257G}. 
Specifically, it is unclear whether AGN activity in these low-mass systems is triggered by external interactions with neighboring massive galaxies \citep{Satyapal2014ApJ...784..113S,Martini2018MNRAS.476.2801M}, or purely by internal, secular processes. 
Analyzing the spatial distribution of these galaxies provides an independent method to robustly test whether active and normal dwarf galaxies reside in distinctly different local environments \citep[e.g.,][]{Li2006MNRAS.368...21L,Yang2018MNRAS.480.1022Y}. 
To determine if proximity to massive structures preferentially triggers AGN activity in the low-mass regime, we compare the local environments of our X-ray-detected AGNs against a $(M_\ast,z)$-matched comparison sample of field dwarf galaxies \citep[e.g.,][]{Kristensen2020MNRAS.496.2577K,Davis2022MNRAS.511.4109D}.

Following \citetalias{Zou2023ApJ...950..136Z}, we construct a control sample of inactive dwarf galaxies from the mass-complete sample described above. 
For each active dwarf galaxy with a given stellar mass ($\log M_\ast$) and redshift ($z$), we randomly select control galaxies that lie within a window of $\log M^{\mathrm{C}}_\ast = (1\pm0.1)\log M_\ast$ and $z^{\mathrm{C}} = z\pm0.075(1+z)$. 
This matching process yields a total of $\sim 2000$ random control galaxies. The sSFRs of the control sample closely match those of the X-ray-detected AGNs; both samples consist primarily of star-forming galaxies, thereby ensuring our comparison is robust against biases introduced by host-galaxy types that correlate with environmental factors.
To quantify environmental influences, we employ the projected separations to neighboring massive galaxies as an environmental indicator, following the methodologies of \citet{Davis2022MNRAS.511.4109D} and \citetalias{Zou2023ApJ...950..136Z}. First, we measure the dispersion of the photometric redshift estimates, denoted as $\sigma_{\Delta z / (1+z)}$, as a function of $z$ using the DJA photo-$z$ catalog. We find this dispersion remains roughly constant at $\approx 0.015$ across the redshift range $z = 0 - 4$.
Following the redshift-slice definition from \citet{Yang2018MNRAS.480.1022Y} (also see \citealt{Darvish2015ApJ...805..121D,Malavasi2016A&A...585A.116M}), we identify massive neighbors of $M^{\mathrm{MG}}_\ast > 10^{10.5}~M_\odot$ for each dwarf galaxy at redshift $z$ that reside within a window of $z^\mathrm{MG} = z \pm 1.5(1+z)\sigma_{\Delta z / (1+z)}$. We then calculate the projected separations to the first-, third-, and tenth-closest massive galaxies. These distances effectively trace the local environment across scales ranging from $\sim 100$~kpc to $5$~Mpc.
Figure~\ref{fig:local_environment} presents the distributions of these projected separations, alongside their median values and associated $1\sigma$ scatter, which is computed as $1.4826 \times$ the median absolute deviation (MAD). Overall, the X-ray-selected dwarf AGN sample exhibits no statistically significant environmental differences compared to the dwarf-galaxy control sample. 
Although the median values for the tenth-closest separation appear to diverge slightly, this is likely driven by the scatter inherent to small-number statistics.
To rigorously test this, we perform two-sample Kolmogorov-Smirnov (K-S) tests comparing the projected separations of the X-ray AGNs and the control dwarf-galaxy sample relative to the massive galaxies. 

Across all three separation metrics, the two-sample K-S tests do not reject the null hypothesis that the X-ray AGNs and matched control dwarfs are drawn from the same projected-environment distribution ($p > 0.05$). 
As an additional validation, we repeat the analysis using massive galaxies drawn from the larger-area CANDELS catalog, and the conclusion remains unchanged. 
Overall, our results are consistent with previous studies finding no statistically significant evidence that active dwarf galaxies reside in different projected environments from matched inactive dwarfs \citep{Davis2022MNRAS.511.4109D, Siudek2023MNRAS.518..724S, Zou2023ApJ...950..136Z}. 
Because this study covers a relatively small solid angle, with an effective area of $\lesssim$ 120 arcmin$^2$, and because the comparison relies on projected separations within photometric-redshift slices, this test is primarily sensitive to relatively strong environmental differences. 
Subtler differences in group membership, satellite/central status, close-pair history, or merger stage may remain undetected.

\subsubsection{Morphological Information}\label{sec:morphological_information}
\begin{figure*}[ht]
    \centering
    \includegraphics[width=\linewidth]{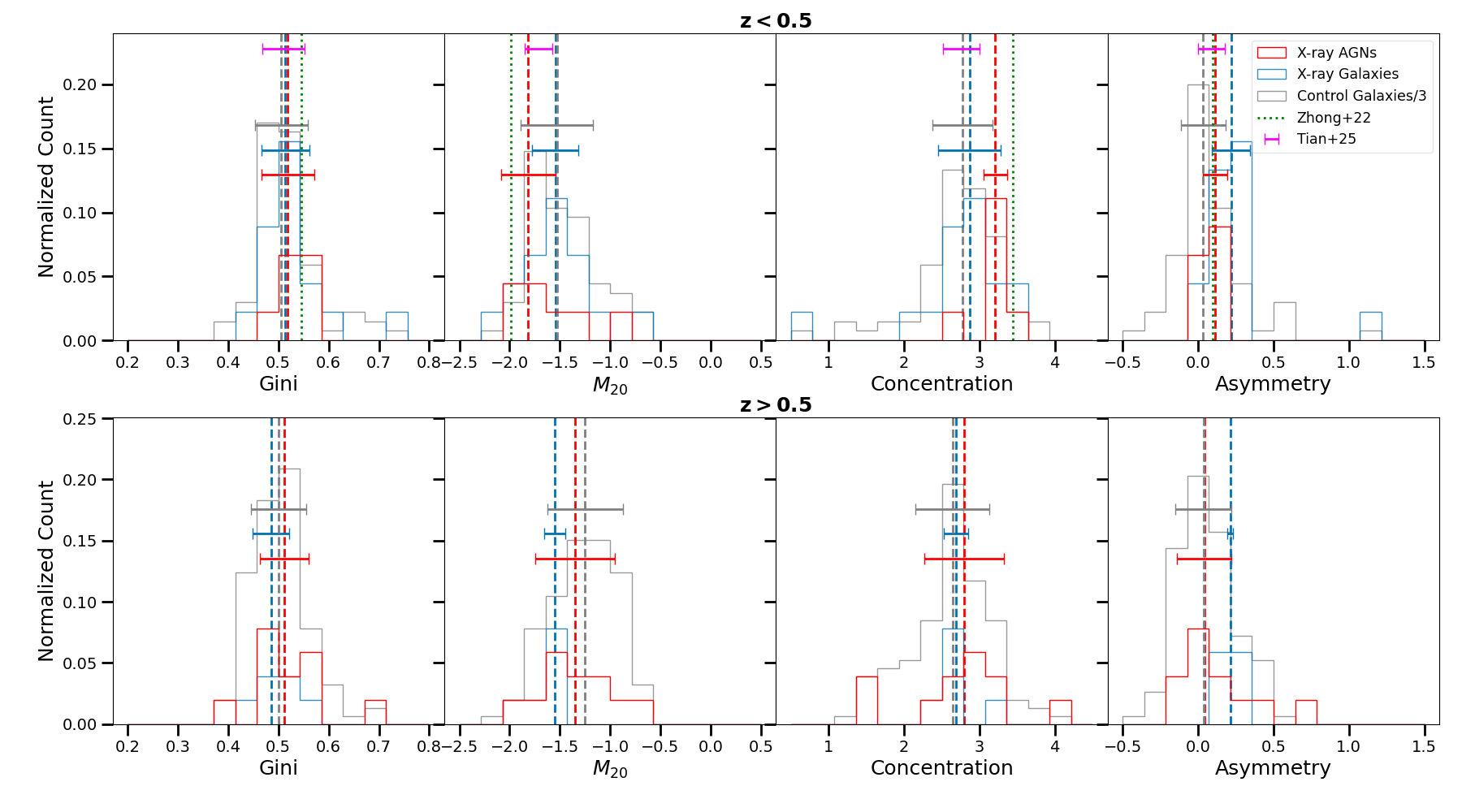}
    \caption{Normalized distributions of non-parametric morphological parameters (Gini, $M_{20}$, Concentration, and Asymmetry) evaluated across two redshift bins: $z < 0.5$ (top row) and $z > 0.5$ (bottom row). 
    The histograms compare the structural properties of X-ray dwarf AGNs (red), X-ray dwarf galaxies (blue), and a mass- and redshift-matched sample of field galaxies (gray, counts are scaled down by a factor of 3 for clarity). 
    The horizontal bars denote the median values and 1$\sigma$ ranges for each respective sample. 
    For comparative context, median values from the recent literature are overlaid in the low-redshift bin: \citet{Zhong2022ApJ...925..157Z} is represented by green dotted vertical lines, and \citet{Tian2025ApJ...986...66T} is represented by magenta markers with error bars. 
    The extensive overlap among the distributions across all four parameters indicates that there is no clear morphological distinction between the active and inactive dwarf galaxies in this sample.}
    \label{fig:cas_opt}
\end{figure*}

We analyze the non-parametric \textit{GMCA} morphological parameters outlined in Section~\ref{sec:gmca}, using the F606W, F814W, or F090W optical bands, depending on filter availability. 
Because these observed filters correspond to different rest-frame wavelengths over $z=$ 0--4, our morphology measurements should be interpreted as observed-frame structural diagnostics rather than uniform rest-frame optical morphologies. 
We mitigate this limitation by comparing the X-ray AGN sample to mass- and redshift-matched field controls measured in the same available imaging.
Therefore, first-order morphological $k$-corrections should affect the active and control samples similarly within each redshift bin.
While redder JWST imaging (e.g., F444W) is available, because the majority of our sample is located at $z < 1$, adopting these filters ensures our structural measurements trace similar rest-frame optical and near-UV wavelengths as the low-redshift comparison samples in the literature \citep[e.g.,][]{Zhong2022ApJ...925..157Z,Tian2025ApJ...986...66T}. This maximizes direct comparability and establishes a consistent baseline for our evolutionary analysis.

Figure~\ref{fig:cas_opt} presents the distributions of Gini, $M_{20}$, concentration, and asymmetry for X-ray AGNs, X-ray galaxies, and matched field dwarfs, separated into $z < 0.5$ and $z > 0.5$ bins. 
The distributions show substantial overlap among all three populations. 
Although the AGN sample may exhibit modest median offsets toward higher concentration and lower asymmetry in some bins, these offsets are small compared with the distribution widths and are not sufficient to define a distinct morphological class. 
We therefore find no compelling evidence that X-ray-selected dwarf AGNs preferentially occupy strongly disturbed or merger-like hosts.
This result argues against recent major mergers being a prerequisite condition for detectable X-ray activity in the present dwarf AGN sample. 
However, the data do not exclude minor interactions, past mergers with long delay times, or stochastic/secular fueling mechanisms operating below the sensitivity of non-parametric morphology diagnostics. 
The relatively low asymmetry values are consistent with a picture in which large-scale disturbances are not the dominant observable signature of AGN triggering in most of these systems \citep[][]{Kaviraj2026MNRAS.545f1918K}, but they do not by themselves prove a secular origin for the accretion.

The X-ray AGNs have median concentration indices of $C = 3.21 \pm 0.60$ at $z < 0.5$ and $C = 2.80 \pm 0.53$ at $z > 0.5$, with median asymmetries of $A \approx$ 0.04--0.11. These values are comparable to those of the mass- and redshift-matched control dwarfs within the observed scatter, indicating no strong enhancement in merger-like morphological features among the AGN hosts.
This result is broadly consistent with previous studies of dwarf AGNs selected at other wavelengths. For example, \citet{Zhong2022ApJ...925..157Z} measured morphological parameters for optical variability-selected AGNs at $z=0$--3 and found that their star-forming host galaxies exhibit asymmetry distributions similar to those of mass- and redshift-matched inactive galaxies. 
Similarly, \citet{Tian2025ApJ...986...66T} analyzed the optical morphologies of a large sample of local AGN-host dwarf galaxies and reported no significant differences in most morphological parameters relative to non-AGN dwarf galaxies. 
Although these studies employed different AGN selection methods and sample definitions, leading to variations in the relative fractions of early- and late-type galaxies, the reported differences are modest. 
As shown in Figure~\ref{fig:cas_opt}, the samples from all three studies occupy similar regions of \textit{GMCA} parameter space. 
Together, these results suggest that the absence of a clear merger-like morphological signature is not unique to our X-ray-selected sample.

Nevertheless, morphology-based merger diagnostics are subject to important limitations. Their sensitivity depends on imaging depth, rest-frame wavelength, spatial resolution, and the time delay between a galaxy interaction and the onset of observable AGN activity \citep[e.g.,][]{Vijarnwannaluk2025ApJ...994..265V}. 
Moreover, while dwarf galaxies generally experience quieter assembly histories and lower overall merger rates than massive galaxies \citep[e.g.,][]{Martin2021MNRAS.500.4937M}, their major merger rates are not dramatically suppressed \citep[e.g.,][]{Casteels2014MNRAS.445.1157C,Rodriguez2015MNRAS.449...49R}.
Recent spectroscopic studies have identified merger-triggered AGNs \citep{Wang2026ApJ..1003..172W} and merger-related dual AGNs \citep{Mezcua2026arXiv260413170M} in individual low-mass galaxies, indicating that mergers can still contribute to SMBH growth and galaxy evolution in at least some dwarf systems.
Therefore, our results should be interpreted as evidence against a dominant role for prominent ongoing major mergers in the triggering of the AGNs studied here, rather than as a definitive exclusion of all externally driven fueling mechanisms. 
AGN hosts may consequently exhibit relatively smooth or centrally concentrated stellar light distributions even if past interactions have influenced their evolution. 

\subsubsection{Nuclear Point-Source Assessment}\label{sec:nuclear_ps}
Most of the X-ray dwarf AGNs do not show apparent unresolved nuclear components in the image cutouts. Consistently, the SED fitting indicates that the AGN component does not dominate the rest-frame UV/optical emission, even for some sources with relatively high X-ray luminosities ($L_\mathrm{X}\sim3\times10^{42}~\mathrm{erg\,s^{-1}}$). 
As an additional check, we perform two-dimensional image decomposition with \texttt{galight}\footnote{https://github.com/dartoon/galight}\citep{Ding2020ApJ...888...37D,Birrer2021JOSS....6.3283B} to assess whether an unresolved nuclear component contributes significantly to the rest-frame UV/optical light of the X-ray AGNs.

For each AGN, we construct an image cutout using the available HST broadband filter that most closely samples the rest-frame UV/optical wavelength range of 2000--6000~\AA. For sources without a sufficiently clear detection in the bluer band, a redder band is used. For most sources, the selected filter falls within the observed-frame wavelength range of 0.4--0.9~$\mu$m. Sources without sufficiently clear imaging for reliable decomposition are excluded from this analysis.
We fit each cutout with a point-source (PS) plus S\'ersic model, placing the PS component at either the galaxy center or the local surface-brightness maximum.

Most of the X-ray AGNs are adequately described by a single S\'ersic profile, and the inclusion of a PS component does not generally yield a significant improvement in the fit.
For the subset of sources with well-constrained decompositions ($\sim60\%$ of the X-ray AGNs), we extract the best-fit PS flux and estimate the PS flux fraction relative to the total flux. 
Most sources have PS flux fractions below $20\%$, with a median value of $\sim8\%$. 
The absence of strong rest-frame UV/optical AGN emission is therefore consistent with its relatively low inferred PS flux fractions, indicating that the host galaxy dominates the rest-frame UV/optical SED. 

One possible explanation for weak rest-frame UV/optical nuclear emission is dust attenuation. 
However, the SED fits do not generally indicate large host-scale color excesses; most sources have low inferred reddening, typically $E(B-V)\sim0$--0.3. 
Therefore, broad-band dust attenuation alone does not appear sufficient to explain the lack of prominent rest-frame UV/optical point sources, although compact nuclear obscuration cannot be ruled out. 
In addition, none of the X-ray-selected dwarf AGNs has a counterpart in the available radio catalogs, based on the cross-matching described in Section~\ref{subsec:agn_classification}. This lack of radio detections suggests that radio-loud or jet-dominated emission is unlikely to be the primary origin of their X-ray excesses.

We further examine the relative strength of the X-ray and UV emission using $\alpha_{\rm OX}$, which characterizes the effective spectral slope between the rest-frame 2500~\AA\ and 2 keV luminosity densities \citep{Tananbaum1979ApJ...234L...9T}:
\begin{equation}
    \alpha_\mathrm{OX}
    = 0.3838 \log \left(
    \frac{L_{\nu,2\,\mathrm{keV}}}{L_{\nu,2500\,\mathrm{\AA}}}
    \right).
\end{equation}
For the X-ray luminosity, we subtract the expected contribution from XRBs and hot gas to estimate the AGN component, and then convert the integrated 2--10~keV luminosity to the monochromatic luminosity density at
2~keV assuming a power-law spectrum with a photon index of $\Gamma=1.8$.

We estimate the rest-frame 2500~\AA\ luminosity density using the nearest available broad-band measurement.
For sources with reliable image decompositions, we multiply the measured broad-band flux by the best-fit PS flux fraction to estimate the AGN contribution.
For sources without reliable PS decompositions, we conservatively attribute the entire broad-band flux to the AGN. The resulting $L_{\nu,2500\,{\mathrm \AA}}$ therefore represents an upper limit, corresponding to a lower limit on $\alpha_{\rm OX}$. We extrapolate from the effective rest-frame wavelength of the selected broad band to 2500~\AA\ assuming an AGN continuum of $L_\nu\propto\nu^{-0.5}$.

The resulting X-ray dwarf AGNs occupy relatively low UV luminosities, typically $\log L_{\nu,2500\,{\rm \AA}}\sim25$--29 $\mathrm{erg\,s^{-1}\,Hz^{-1}}$. 
Their $\alpha_{\rm OX}$ values are typically around $-1.5$ but extend to lower values, lying below the relation measured for luminous unobscured quasars \citep[e.g.,][]{Just2007ApJ...665.1004J}, with a typical offset of $\Delta \alpha_{\rm OX}=\alpha_{\rm OX}-\alpha_{\rm OX,Just+07}\sim-0.37$. 
This offset indicates that the sources are X-ray weak relative to their inferred UV luminosities, or equivalently that the UV luminosities are likely overestimated for the AGN component. 
However, low-luminosity AGNs and dwarf AGNs often show substantial scatter in $\alpha_\mathrm{OX}$, partly because the nuclear UV continuum is difficult to isolate from host-galaxy emission \citep[e.g.,][]{Dong2012ApJ...755..167D}. 
The values found here overlap with the broad range reported for low-luminosity and dwarf AGNs in the local universe \citep[e.g.,][]{Yuan2014ApJ...782...55Y,Plotkin2016ApJ...825..139P,Baldassare2017ApJ...836...20B,Wasleske2023AJ....166...64W}, but the lower $\alpha_{\rm OX}$ tail may be enhanced by residual host-galaxy contamination to the UV/optical luminosities. 
For sources without reliable PS decompositions, our adopted $L_{\nu,2500\,{\rm \AA}}$ values should therefore be interpreted as conservative upper limits to the AGN UV luminosity, implying lower limits on $\alpha_{\rm OX}$. 
Thus, the comparison with both luminous quasars and low-luminosity AGNs supports the interpretation that the rest-frame UV/optical emission in many of our sources may be strongly diluted by the host galaxy.

The absence of rest-frame UV/optical nuclear point sources is also consistent with the moderate accretion rates inferred for the sample. 
Because AGN luminosity correlates with host-galaxy stellar mass, a more informative measure of black-hole growth is the specific accretion rate, $\lambda=L_{\rm X}/M_\ast$ \citep[e.g.,][]{Aird2018MNRAS.474.1225A,Yang2018MNRAS.475.1887Y,Zou2024ApJ...964..183Z}. 
Most of our sources have $\log\lambda\lesssim33$, corresponding approximately to $\lambda_{\rm Edd}\sim10^{-3}$--$10^{-1}$ with commonly adopted bolometric corrections and canonical $M_{\rm BH}$--$M_\ast$ scaling relations. 
Thus, the sample is generally consistent with moderate or low-level accretion, but not necessarily with a purely radiatively inefficient accretion regime.

Several of the most X-ray-luminous dwarf AGNs are also located at $z \gtrsim 1$, where additional observational effects may further suppress detectable nuclear signatures. 
Galaxies become increasingly compact toward higher redshift \citep[e.g.,][]{vanDerWel2014ApJ...788...28V}, reducing the spatial contrast between the nucleus and the host. In addition, both observations and theoretical models suggest that AGN obscuration becomes more prevalent in the early universe as gas fractions increase and galaxies become more centrally concentrated \citep[e.g.,][]{Gilli2022A&A...666A..17G}. 

The combination of intrinsically weak UV/optical AGN continua, host-galaxy dilution, moderate accretion rates, possible compact nuclear obscuration, and the compact structures of high-redshift host galaxies likely contribute to the absence of rest-frame UV/optical nuclear point sources in many X-ray-selected dwarf AGNs.

\subsection{AGN Accretion Properties, AGN Fraction, Long-term BH Accretion}\label{sec:agn_fraction_bhar}
\begin{figure*}[ht]
    \centering
    \includegraphics[width=\linewidth]{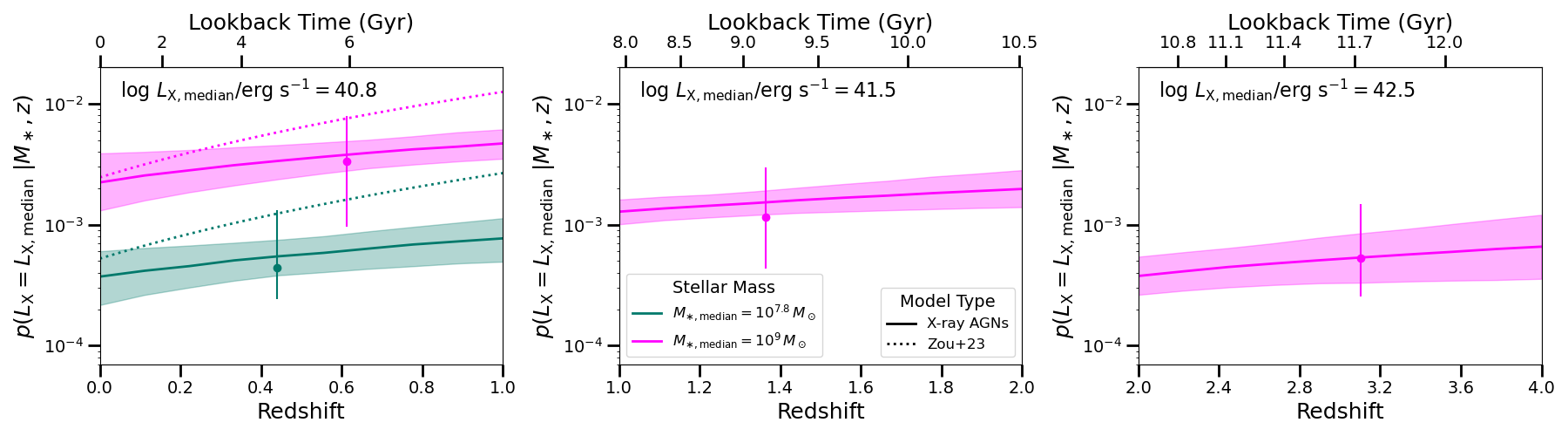}
    \caption{
    \textit{Left, middle, and right panels:} Accretion-rate distribution functions in three redshift bins, evaluated at the median $L_\mathrm{X}$ of each $M_\ast-z$ bin. Two representative stellar masses are shown: $M_{\ast,\mathrm{median}} = 10^{7.8}~M_\odot$ (dark cyan) and $M_{\ast,\mathrm{median}} = 10^{9}~M_\odot$ (magenta). 
    Solid curves denote the medians of the posterior sampling distributions, with shaded regions indicating the $1\sigma$ uncertainties derived from the sampling percentiles.
    Dotted curves represent the results from \citetalias{Zou2023ApJ...950..136Z}, scaled to the $z$--$M_\ast$--$L_\mathrm{X}$ coverage of the current sample. 
    Filled circles with error bars represent binned measurements, calculated by incorporating the expected AGN counts given the survey depth and median luminosity in each bin. Constraints are strongest at $z < 1$, while only the higher-mass bin ($M_\ast \gtrsim 10^{9}~M_\odot$) is populated at $z > 1$.}
    \label{fig:pLx}
\end{figure*}
We fit the X-ray accretion-rate distribution described in Section~\ref{sec:accretion_modeling} to the 27 X-ray AGNs in our mass-complete dwarf-galaxy sample that satisfy $L_\mathrm{X}=10^{40}-10^{43}\mathrm{~erg~s^{-1}}$. 
The best-fit parameters are reported in Table~\ref{tab:pLx_params}. 
To compare the continuous parametric model with the discrete observations, we compute binned empirical estimates of the probability density following \citet{Aird2012ApJ...746...90A}. 
The parent galaxy and AGN samples are partitioned into predefined redshift ($z \in [0, 1, 2, 4]$) and stellar mass ($M_\ast \in [10^7, 10^{8.25}, 3 \times 10^9] \, M_\odot$) bins. 
Within each bin, after applying the mass-completeness limits to the parent sample, we evaluate the expected number of AGNs, $N_\mathrm{exp}$, using Equations~\ref{eq:pdet_fx} and \ref{eq:n_mdl} for a given parent galaxy population. 
The empirical binned probability density is then derived by scaling the model probability density evaluated at the median $z$--$M_\ast$--$L_\mathrm{X}$ values (shown in Figure~\ref{fig:pLx}) of the observed AGNs by the ratio $N_\mathrm{obs} / N_\mathrm{exp}$. 
Uncertainties are computed using the small-number Poisson confidence limits of \citet{Gehrels1986ApJ...303..336G}. 
The binned estimates are in good agreement with the best-fit model.

\begin{deluxetable}{lcc}
\tablecaption{Best-fit Parameters of Equation~\ref{eq:pLx}. \label{tab:pLx_params}}
\tablecolumns{3}
\tablewidth{0pt}
\tablehead{
\colhead{Parameter} & \colhead{X-ray dwarf AGNs} & \colhead{Joint dwarf AGNs}  \\
\colhead{} & \colhead{(Median $\pm 1\sigma$)} & \colhead{(Median $\pm 1\sigma$)} }
\startdata
$\log A$   & $-2.78^{+0.47}_{-0.48}$ & $-2.98^{+0.19}_{-0.18}$ \\
$\gamma_M$ & $0.67^{+0.13}_{-0.15}$  & $0.44^{+0.09}_{-0.09}$ \\
$\gamma_L$ & $-0.75^{+0.14}_{-0.15}$ & $-0.58^{+0.06}_{-0.05}$ \\
$\gamma_z$ & $1.1^{+0.9}_{-1.0}$     & $1.1^{+0.3}_{-0.4}$ \\
\enddata
\tablecomments{The median values and 1$\sigma$ credible intervals are derived from the marginalized MCMC posteriors.}
\end{deluxetable}

Our derived model indicates that the incidence of X-ray AGN activity depends on stellar mass, X-ray luminosity, and redshift.
For the X-ray-selected dwarf AGN sample, the best-fit normalization is $\log A=-2.78^{+0.47}_{-0.48}$.
We find a positive stellar-mass dependence ($\gamma_M=0.67^{+0.13}_{-0.15}$), indicating that the probability density $p(L_\mathrm{X}\,|\,M_\ast,z)$ increases toward higher-mass dwarf galaxies.
This behavior is qualitatively consistent with previous studies of AGN populations in more massive hosts \citep[e.g.,][]{Aird2018MNRAS.474.1225A,Zou2024ApJ...964..183Z}, although our measurements show that the mass dependence extends toward the dwarf-galaxy regime.
The luminosity dependence is characterized by $\gamma_L=-0.75^{+0.14}_{-0.15}$, implying that high-luminosity AGNs are substantially less common than their lower-luminosity counterparts, consistent with the declining shape of AGN luminosity distributions found in previous work.
We also infer positive redshift evolution, with $\gamma_z=1.1^{+0.9}_{-1.0}$. 
This result indicates that the incidence of AGN activity rises toward earlier cosmic epochs, suggesting that BH growth in dwarf galaxies was substantially more common at $z>1$ than in the local universe. Although the uncertainty on $\gamma_z$ remains non-negligible, the posterior distribution strongly favors positive evolution, providing evidence that AGN activity in dwarf galaxies tracks the broader rise of cosmic accretion activity toward cosmic noon.

Integrating the $p(L_\mathrm{X}\,|\,M_\ast,z)$ over the adopted luminosity range allows us to estimate the observed AGN fraction as a function of stellar mass and redshift. Our results are best constrained at $z<1$, where the majority of X-ray AGNs in our sample are detected. 
At $z\approx0$, the observed AGN fraction ($f_\mathrm{AGN}$ for $L_\mathrm{X} = 10^{40}$--$10^{43} \, \mathrm{erg\,s^{-1}}$) is of order $10^{-3}$ for $M_\ast=10^{7.8}\,M_\odot$ dwarf galaxies and several $\times 10^{-3}$ for $M_\ast=10^9\,M_\odot$ dwarf galaxies.
The positive redshift dependence inferred from the fit predicts a gradual increase in AGN incidence toward higher redshift.
For $M_\ast=10^9\,M_\odot$, the model reaches $f_\mathrm{AGN}\sim10^{-2}$ by $z\sim2$ and a few percent by $z\sim4$, whereas the lower-mass curve is only shown to the redshift where the parent galaxy sample remains mass complete.
We emphasize that these high-redshift estimates rely partially on extrapolating the fitted luminosity dependence below the direct X-ray detection threshold and should therefore be interpreted as model-dependent predictions.

\begin{figure}
    \centering
    \includegraphics[width=\linewidth]{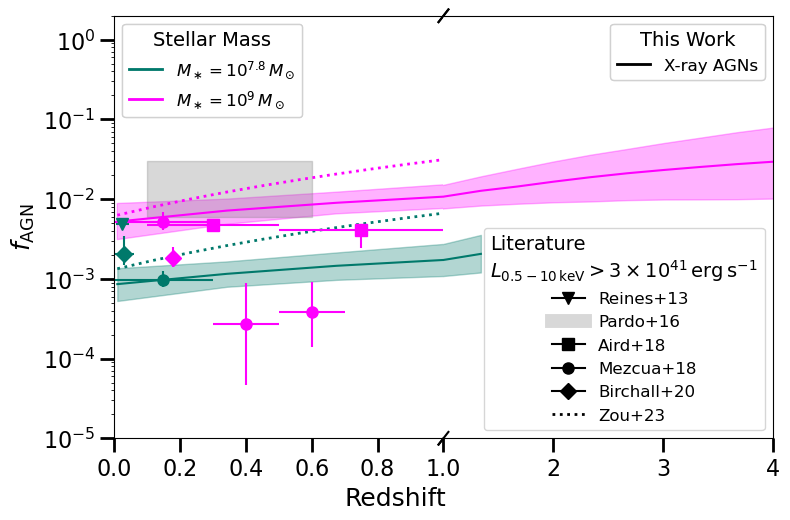}
    \caption{
    Evolution of the AGN fraction ($f_\mathrm{AGN}$) as a function of redshift. 
    We compute the AGN fraction by integrating the posterior probability density distributions for sources with $L_\mathrm{X}=10^{40}$--$10^{43}~\mathrm{erg~s^{-1}}$. 
    The mass dependence is illustrated by the dark cyan ($M_\ast = 10^{7.8}~M_\odot$) and magenta ($M_\ast = 10^{9}~M_\odot$) solid curves, with the shaded regions indicating the $1\sigma$ uncertainties. 
    These curves are truncated at higher redshifts where the respective stellar mass falls below our observational mass completeness limit (as shown in Figure~\ref{fig:prop1}), ensuring that the plotted fractions rely on robust measurements.
    For comparison, we overplot observational constraints from several $z < 1$ dwarf AGN studies, color-coded to correspond to our adopted stellar-mass bins. 
    Where necessary, the literature values are derived by summing the contributions across all available luminosity bins. The comparison sample includes \citet{Reines2013ApJ...775..116R} (downward triangles), \citet{Pardo2016ApJ...831..203P} (gray shaded region), \citet{Aird2018MNRAS.474.1225A} (squares), \citet{Mezcua2018MNRAS.478.2576M} (circles), \citet{Birchall2020MNRAS.492.2268B} (diamonds), and \citetalias{Zou2023ApJ...950..136Z} (dotted curves). 
    A broken $x$-axis is employed to expand the low-redshift ($z \le 1$) regime for clearer visualization of the literature constraints, while the $z > 1$ range is compressed.}
    \label{fig:agn_fraction}
\end{figure}

Notably, however, our results for the lowest-mass regime ($M_\ast = 10^{7.8}~M_\odot$) are only depicted up to $z \approx 1.2$. 
Beyond this redshift, these galaxies fall below the mass-completeness limit threshold of the survey, rendering us unable to confidently constrain the active fraction at the extreme low-mass end during these earlier epochs. 
For the same reason, the lowest-mass curves in the accretion-rate distribution, $f_\mathrm{AGN}$, and $\overline{ \mathrm{BHAR}}$ panels are terminated at lower redshift than the $M_\ast = 10^9~M_\odot$ curves: beyond these redshifts, the parent dwarf-galaxy sample is no longer mass complete, and extrapolating the fitted distribution would not be directly constrained by the observed galaxy population.
We note that the comparison with \citet{Mezcua2018MNRAS.478.2576M} in Figure~\ref{fig:agn_fraction} should be interpreted with this luminosity-bin coverage in mind. For consistency with our luminosity-integrated $f_\mathrm{AGN}$, the literature values are constructed by summing over the available luminosity bins at fixed stellar mass. 
However, in the higher-redshift bins of \citet{Mezcua2018MNRAS.478.2576M}, measurements are available only for the higher-luminosity intervals. 
The apparently lower $f_\mathrm{AGN}$ at higher redshift in that comparison therefore reflects incomplete coverage of the lower-luminosity AGN population, rather than evidence for weaker cosmic evolution.

\begin{figure}
    \centering
    \includegraphics[width=\linewidth]{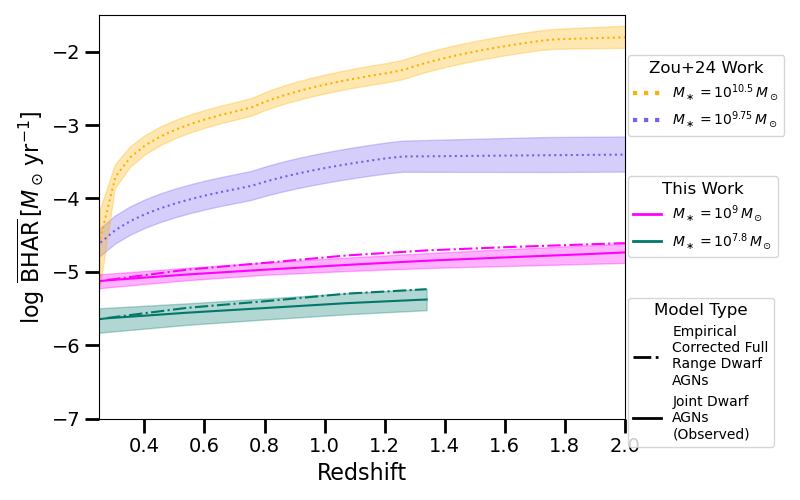}
    \caption{
     Evolution of the $\overline{\mathrm{BHAR}}$ as a function of redshift for different stellar-mass regimes. 
    The solid dark-cyan ($M_\ast=10^{7.8}~M_\odot$) and magenta ($M_\ast=10^{9}~M_\odot$) curves show the joint dwarf-galaxy estimates obtained by combining the X-ray dwarf AGNs in this work with the dwarf AGNs from \citetalias{Zou2023ApJ...950..136Z}. 
    These curves represent the observed dwarf-AGN contribution at $L_\mathrm{X}=10^{40}$--$10^{44}~\mathrm{erg~s^{-1}}$, and the shaded regions indicate the $1\sigma$ uncertainty around the median. 
    The dash-dotted curves show the empirically corrected dwarf-galaxy estimates, where a redshift-dependent fractional correction is applied to account for the possible contribution from the unobserved high-luminosity tail at $L_{\rm X}>10^{44}~\mathrm{erg~s^{-1}}$. 
    The correction is calibrated from the ratio between the $\overline{\mathrm{BHAR}}$ predicted by the \citet{Zou2024ApJ...964..183Z} relation over $L_{\rm X}=10^{44}$--$10^{46}~\mathrm{erg~s^{-1}}$ and that over $L_{\rm X}=10^{41}$--$10^{44}~\mathrm{erg~s^{-1}}$, evaluated at $M_{\ast,\rm ref}=10^{9.75}~M_\odot$. 
    For comparison, the dotted orange and purple curves show the massive-galaxy relations from \citet{Zou2024ApJ...964..183Z} for $M_\ast=10^{10.5}~M_\odot$ and $M_\ast=10^{9.75}~M_\odot$, respectively, integrated over $L_\mathrm{X}=10^{41}$--$10^{46}~\mathrm{erg~s^{-1}}$.
    }
\label{fig:bhar}
\end{figure}

Our CDF dwarf sample does not directly constrain the rare high-luminosity AGN population at $L_\mathrm{X}>10^{43}\,\mathrm{erg\,s^{-1}}$, which is preferentially hosted by more massive galaxies. 
In dwarf galaxies, such luminosities would generally require high Eddington ratios and/or relatively massive BHs for their host stellar masses.
These episodes are expected to be rare, as the shallow gravitational potentials of dwarf galaxies are unlikely to sustain prolonged episodes of high-rate BH accretion. 
Because $\overline{\mathrm{BHAR}}\propto L_\mathrm{max}^{1+\gamma_{\mathrm{L}}}\sim L_\mathrm{max}^{0.25}$, the population-averaged accretion rate is sensitive to the adopted upper luminosity limit. Consequently, extrapolating the luminosity distribution beyond the directly observed range could bias the inferred total accretion budget.

To maintain observationally motivated constraints, we restrict our primary analysis to X-ray dwarf AGNs with $L_\mathrm{X}=10^{40}-10^{43}\,\mathrm{erg\,s^{-1}}$. 
Estimating the full intrinsic $\overline{\mathrm{BHAR}}$ requires accounting for the contribution of the unobserved high-luminosity tail at $L_\mathrm{X}>10^{43}\,\mathrm{erg\,s^{-1}}$, which is not directly constrained by our X-ray dwarf AGN sample. 
To assess the impact of this high-luminosity contribution, we repeat the fit after including the dwarf AGNs from \citetalias{Zou2023ApJ...950..136Z}, which extend the dwarf-AGN luminosity coverage to $L_\mathrm{X}\lesssim10^{44}\,\mathrm{erg\,s^{-1}}$ at $z<1$. 
The resulting joint-sample estimates, integrated over $L_\mathrm{X}=10^{40}$--$10^{44}\,\mathrm{erg\,s^{-1}}$, are shown as the solid curves in Figure~\ref{fig:bhar}. 
The joint fit produces only modest changes in the shape parameters of $p(L_\mathrm{X}\mid M_\ast,z)$, as listed in Table~\ref{tab:pLx_params}. 
For a $10^9\,M_\odot$ dwarf galaxy, $\overline{\mathrm{BHAR}}$ increases gradually with redshift, consistent with the positive redshift dependence inferred from our parameterized $p(L_\mathrm{X}\mid M_\ast,z)$ distribution. 
At fixed redshift, the lower-mass $M_\ast=10^{7.8}\,M_\odot$ dwarfs have systematically lower $\overline{\mathrm{BHAR}}$, in agreement with the positive stellar-mass dependence implied by the accretion-rate distribution.

Compared with more massive galaxies, dwarf galaxies show substantially lower average BH growth rates. 
Nevertheless, they follow a qualitatively similar evolutionary pattern: $\overline{\mathrm{BHAR}}$ increases with both stellar mass and redshift. 
Even the joint dwarf-galaxy estimates remain well below the massive-galaxy relations from \citet{Zou2024ApJ...964..183Z}, indicating that, although BH growth in dwarfs becomes more active toward earlier cosmic epochs, its normalization is still significantly lower than in more massive systems. 
These results indicate that the stellar-mass dependence of BH growth established in massive galaxies extends into the dwarf-galaxy regime \citepalias{Zou2023ApJ...950..136Z}.

Although the joint dwarf sample improves the observational coverage relative to the CDF-only sample, it still does not directly constrain the rare high-luminosity tail at $L_{\rm X}>10^{44}~\mathrm{erg~s^{-1}}$. 
To estimate the possible contribution from the unobserved high-luminosity tail, we apply an empirical luminosity-range correction based on the massive-galaxy accretion-rate distribution from \citet{Zou2024ApJ...964..183Z}. 
For each redshift, we first evaluate the \citet{Zou2024ApJ...964..183Z} relation at a reference stellar mass of $M_\ast=10^{9.75}~M_\odot$ and compute the $\overline{\mathrm{BHAR}}$ contribution over the luminosity range directly sampled by the massive-galaxy calibration, $L_{\rm X}=10^{41}$--$10^{44}~{\rm erg~s^{-1}}$, and over the high-luminosity tail, $L_{\rm X}=10^{44}$--$10^{46}~{\rm erg~s^{-1}}$.
We then define the fractional high-luminosity contribution as  
\begin{align}
R_{L_{\rm X}}(z) =\frac{\overline{\mathrm{BHAR}}_{\rm tail}(L_\mathrm{X}=10^{44}\mathrm{-}10^{46}~\mathrm{erg\,s^{-1}})}
{\overline{\mathrm{BHAR}}_{\rm sampled}(L_\mathrm{X}=10^{41}\mathrm{-}10^{44}~\mathrm{erg\,s^{-1}})}.
\end{align}
The empirically corrected dwarf-galaxy $\overline{\mathrm{BHAR}}$ is obtained by applying this fractional boost to the joint dwarf-AGN estimate:
\begin{align}
\log \overline{\mathrm{BHAR}}_{\rm corr} = \log \overline{\mathrm{BHAR}}_{\rm joint} + \log\left[1+R_{L_{\rm X}}(z)\right].
\end{align}
Here $\overline{\mathrm{BHAR}}_{\rm joint}$ is computed from the joint dwarf-AGN fit over 
$L_{\rm X}=10^{40}$--$10^{44}~\mathrm{erg~s^{-1}}$.

This correction should be interpreted as an empirical estimate of the missing high-luminosity contribution rather than a direct measurement of the intrinsic dwarf-galaxy luminosity function at $L_{\rm X}>10^{44}~\mathrm{erg~s^{-1}}$. 
It assumes that the fractional contribution of the high-luminosity tail, relative to the $L_{\rm X}=10^{41}$--$10^{44}\,\mathrm{erg\,s^{-1}}$ range used in the empirical calibration, can be approximated using the better-constrained massive-galaxy relation of \citet{Zou2024ApJ...964..183Z}. 
The corrected estimates, shown as dash-dotted curves in Figure~\ref{fig:bhar}, should therefore be regarded as extrapolated, full-range estimates for dwarf galaxies. 
The correction increases the inferred $\overline{\mathrm{BHAR}}$ by approximately 20--40\% relative to the joint observed estimates, although the corrected dwarf-galaxy values remain well below the corresponding massive-galaxy relations.
This result reinforces the conclusion that dwarf galaxies host lower average BH growth rates than more massive systems, while still exhibiting enhanced accretion activity toward earlier cosmic epochs.

Several limitations should be kept in mind when interpreting these results. 
The physical interpretation of BH growth in dwarf galaxies remains subject to systematic uncertainties in the adopted bolometric corrections and BH--galaxy scaling relations. 
Dwarf galaxies are known to exhibit substantial scatter and possible systematic offsets from canonical $M_\mathrm{BH}$--$M_\ast$ relations \citep[e.g.,][]{Reines2015ApJ...813...82R,Mezcua2017IJMPD..2630021M,Schutte2019ApJ...887..245S}, while the bolometric corrections commonly applied to AGNs may not be directly applicable to low-mass BHs \citep[e.g.,][]{Cann2019ApJ...870L...2C,Duras2020A&A...636A..73D,Zou2023ApJ...950..136Z,Pucha2025ApJ...982...10P}. 
Consequently, quantities such as $\overline{\mathrm{BHAR}}$ should be interpreted with caution, particularly when comparing dwarf galaxies with more massive systems.

\section{Summary and Future Work}
In this paper, we have presented a comprehensive population study of X-ray-detected dwarf galaxies ($M_* < 3 \times 10^9 M_\odot$) in the Chandra Deep Fields North and South. 
By integrating ultra-deep Chandra legacy catalogs with JWST and HST imaging from the DAWN JWST Archive, we performed rigorous SED fitting to decouple AGN contributions from host-galaxy stellar emission. 
Our final host-galaxy mass-complete X-ray-detected sample consists of 56 sources, comprising 31 X-ray AGNs and 25 X-ray galaxies not satisfying our conservative AGN criteria. 
We investigated their local environments, structural morphologies, and the cosmic evolution of their accretion activities. 
Because the analysis is based on the small-area Chandra Deep Fields, the highest-redshift constraints remain sensitive to small-number statistics and cosmic variance. 
Our main conclusions are as follows:
\begin{enumerate}
    \item X-ray Hardness Ratio: X-ray sources at $z < 2$ are predominantly unobscured or only mildly obscured, consistent with local universe findings. At higher redshifts, the stacking analysis suggests a possible emergence of a more obscured dwarf AGN population. See Section~\ref{sec:hardness_ratio}.
    \item Environmental Independence: We found no statistically significant difference between the local environments of X-ray AGNs and a mass- and redshift-matched control sample of field dwarf galaxies. 
    Two-sample Kolmogorov-Smirnov tests comparing projected separations to neighboring massive galaxies yielded $p > 0.05$ across all spatial scales, indicating that we find no statistically significant evidence for systematically different projected environments between the active dwarf galaxies and their matched inactive counterparts. See Section~\ref{sec:local_host_environments}.
    \item Morphology and triggering: Our non-parametric \textit{GMCA} morphological analysis reveals extensive overlap in the structural properties of X-ray dwarf AGNs, X-ray dwarf galaxies, and mass- and redshift-matched field galaxies. The lack of a clear morphological distinction indicates that the active hosts do not preferentially exhibit the highly asymmetric or disturbed structures expected for ongoing major mergers, consistent with recent studies of dwarf AGN hosts \citep[e.g.,][]{Zhong2022ApJ...925..157Z,Tian2025ApJ...986...66T}. This argues against large-scale structural disruption being a dominant or necessary triggering mechanism for the X-ray-selected dwarf AGNs in this sample, although minor interactions, delayed merger-triggered activity, and stochastic or secular fueling remain viable. See Section~\ref{sec:morphological_information}.
    \item Weak Nuclear Point-source Contribution: Most X-ray-selected dwarf AGNs do not show prominent unresolved nuclear point sources in the rest-frame UV/optical imaging. Image-decomposition analysis indicates that a single S\'ersic profile generally provides an adequate description, with only modest PSF contributions where meaningful constraints can be obtained. This weak nuclear component is consistent with the relatively low inferred AGN UV contribution suggested by the $\alpha_{\rm OX}$ analysis, and may reflect a combination of host-galaxy dilution and obscuration. See Section~\ref{sec:nuclear_ps}.
    \item Cosmic Evolution of Black Hole Accretion: By modeling the specific accretion rate distribution, $p(L_\mathrm{X} \,|\, M_\ast, z)$, we infer a mild cosmic evolution in dwarf AGN activity. 
    For a representative $10^9~M_\odot$ dwarf galaxy, the model-integrated active fraction over $L_\mathrm{X}=10^{40}$--$10^{43}~\mathrm{erg~s^{-1}}$ rises gradually with redshift, reaching the $\sim 10^{-2}$ level by $z \sim $2--4, depending on the adopted luminosity limits. 
    The corresponding $\overline{\mathrm{BHAR}}$ also increases toward higher redshift and is systematically higher in more massive dwarf galaxies. To better account for the high-luminosity contribution, we combine our X-ray dwarf AGNs from the CDFs with the dwarf AGNs from \citetalias{Zou2023ApJ...950..136Z}, extending the observed dwarf-AGN luminosity coverage to $L_\mathrm{X}\lesssim10^{44}~\mathrm{erg~s^{-1}}$. 
    We further apply an empirical luminosity-range correction, calibrated from the massive-galaxy relation of \citet{Zou2024ApJ...964..183Z}, to estimate the possible contribution from the unobserved $L_\mathrm{X}>10^{44}~\mathrm{erg~s^{-1}}$ tail. 
    The corrected $\overline{\mathrm{BHAR}}$ remains substantially below that of more massive galaxies, but shows a qualitatively similar trend of increasing BH growth toward earlier cosmic epochs. 
    See Section~\ref{sec:agn_fraction_bhar}.
    \end{enumerate}

In addition to the unprecedented depth of the CDFs and the extensive multiwavelength coverage available in GOODS-S and GOODS-N, this study also highlights several opportunities for further progress in understanding BH growth in dwarf galaxies. 
A key challenge is to improve the separation of weak AGN emission from stellar light, dust attenuation, and star formation. 
Although JWST/NIRCam achieves an angular resolution of $\sim$0.05--0.15$^{\prime\prime}$, corresponding to physical scales of $\sim$0.1--1.3 kpc at $z\sim$0.1--2, spatial decomposition remains challenging for sources with compact hosts with weak nuclear-to-host contrast, consistent with the small nuclear point-source contributions found in our sample.
In addition, larger samples at $z>$2 and wider survey volumes will enable stronger environmental constraints and a more complete assessment of the trends identified here. 
Extending similar analyses to wider-area fields with deep X-ray and JWST observations, such as COSMOS-Web, will therefore be important for expanding the statistical census of dwarf AGNs and testing how representative the CDF results are across different cosmic environments.

Current facilities already provide unique access to the most luminous accreting BHs in low-mass galaxies at high redshift, and next-generation X-ray observatories will extend this view to fainter and more typical systems. 
For example, the Lynx mission aims at delivering an on-axis angular resolution of $\sim0.5^{\prime\prime}$ while maintaining a resolution of $<1^{\prime\prime}$ within a $10^{\prime}$ off-axis radius, together with a substantially greater collecting area than Chandra and a point-source sensitivity reaching $\sim10^{-19}~\mathrm{erg~cm^{-2}~s^{-1}}$ in its deepest surveys \citep{Gaskin2019JATIS...5b1001G}.
Maintaining a narrow PSF across the field would address a limitation of Chandra, whose PSF degrades significantly at larger off-axis angles, particularly affecting single-field surveys such as the CDF-N and CDF-S used in this work.
These capabilities would enable the detection of substantially fainter accreting BHs while mitigating source confusion across deep extragalactic fields.
Combined with deep multiwavelength observations, such facilities will enable more complete measurements of BH accretion in dwarf galaxies across cosmic time and provide stronger empirical constraints on the formation and early growth of SMBH seeds.

\begin{acknowledgments}
WLT, WNB, ZY, and BZ acknowledge support from NSF grant AST-2407089 and the Penn State Eberly Endowment. Some of the data products presented herein were retrieved from the Dawn JWST Archive (DJA). DJA is an initiative of the Cosmic Dawn Center (DAWN), which is funded by the Danish National Research Foundation under grant DNRF140.
\end{acknowledgments}

\clearpage

\appendix
\section{\texttt{CIGALE} Parameter}
Table~\ref{tab:cigale} summarizes the \texttt{CIGALE} modules and parameter grids adopted in the SED fitting described in Section~\ref{sec:sed_fitting_dwarf_selection}. 
\FloatBarrier
\setcounter{table}{0}
\renewcommand{\thetable}{\thesection\arabic{table}}
\begin{deluxetable}{lcl}
\tabletypesize{\footnotesize}
\tablecaption{CIGALE Parameters and Grid Settings\label{tab:cigale}}
\tablewidth{0pt}
\tablehead{
\colhead{Parameter} & \colhead{Value}& \colhead{Description}
}
\startdata
\multicolumn{3}{c}{\textbf{Star Formation History: sfhdelayed}} \\[2pt]
$\tau_{\mathrm{main}}$ & 100, 500, 1000, 5000 Myr & $e$-folding time of main stellar population \\
age$_\mathrm{main}$ & 1000, 2000, 5000 Myr & Age of the main stellar population \\
\hline
\multicolumn{3}{c}{\textbf{Single-age Stellar Population Model: bc03 \citep{Bruzual2003MNRAS.344.1000B}}} \\[2pt]
imf & 1 & \citet{Chabrier2003PASP..115..763C} initial mass function \\
metallicity & 0.0001,0.0004,0.004,0.008,0.02 & Range of stellar population in solar metallicity \\
separation\_age & 10.0 Myr & Old/young stellar age separation \\
\hline
\multicolumn{3}{c}{\textbf{Nebular Model: nebular}}\\ [2pt]
logU & -2 & Ionization parameter \\
zgas & 0.004, 0.008, 0.014, 0.033 & Range for gas metallicity in solar metallicity\\
\hline
\multicolumn{3}{c}{\textbf{Dust Attenuation: dustatt\_modified\_starburst (Modified \citet{Calzetti2000ApJ...533..682C} Attenuation Law)}} \\[2pt]
E\_BV\_lines & 0, 0.1, 0.3, 0.6, 1 & Color excess of the nebular lines  \\
E\_BV\_factor & 0.44 & Stellar $E(B{-}V)$ conversion factor \\
Ext\_law\_emission\_lines & 1, 2 & 1: Milky-Way, 2: LMC\\
\hline
\multicolumn{3}{c}{\textbf{Dust Emission: dl2014 \citep{Draine2014ApJ...780..172D}}} \\[2pt]
qpah & 2.5 & Mass fraction of PAH\\
umin &1.0& Minimum radiation field\\
alpha & 2.0& Power-law slope $\frac{dU}{dM}$ proportional to $U^{-\alpha}$\\
gamma &0.1& Fraction illuminated from $U_\mathrm{min}$ to $U_\mathrm{max}$\\
\hline
\multicolumn{3}{c}{\textbf{AGN Model: skirtor2016 \citep{Stelevski2016MNRAS.458.2288S}}} \\[2pt]
i & 10, 30, 50, 70 & Viewing angle (degree) with respect to the AGN\\
fracAGN & 0, 0.1, 0.25, 0.5, 0.6, 0.7, 0.8, 0.9 & Fraction of AGN dust emission to the IR total luminosity\\
EBV & 0, 0.025, 0.05, 0.075, 0.1, 0.125, 0.15, 0.175 & $E(B-V)$ for the color excess in the polar direction\\
\hline
\multicolumn{3}{c}{\textbf{X-ray Model: yang20 \citep{Yang2020MNRAS.491..740Y,Yang2022ApJ...927..192Y}}} \\[2pt]
gam & 1.8 & Photon index of the AGN intrinsic X-ray spectrum.\\
alpha\_ox & -2.0,-1.75, -1.5,-1.25, -1.0 & Power-law slope connecting $L_\nu$ at rest-frame 2500~\AA~and 2 keV\\
\hline
\enddata
\end{deluxetable}

\clearpage
\onecolumngrid
\section{X-ray Dwarf Catalog}
Table~\ref{tab:catalog} lists the X-ray-detected dwarf galaxies in the CDFs, together with their source identifiers, coordinates, redshifts, X-ray luminosities, \texttt{CIGALE}-derived physical properties, and AGN classifications. 
The detailed sample-selection and AGN-classification procedures are described in Sections~\ref{sec:sed_fitting_dwarf_selection} and \ref{subsec:agn_classification}, respectively.
\setcounter{table}{0}
\renewcommand{\thetable}{\thesection\arabic{table}}
\startlongtable
\begin{deluxetable}{lccccccccc}
\tabletypesize{\footnotesize}
\tablecaption{X-ray Dwarf AGNs and Galaxies in CDFs. \label{tab:catalog}}
\tablewidth{0pt}
\tablehead{
\colhead{XID} & \colhead{NID} & \colhead{RA}& \colhead{DEC} & \colhead{z} & \colhead{ztype} & \colhead{$\log L_\mathrm{X}$} & \colhead{$\log M_\ast$} &\colhead{$\log \mathrm{SFR}$}   &\colhead{Type} \\
& & \colhead{deg} & \colhead{deg} & & & \colhead{$\mathrm{erg\,s^{-1}}$} & \colhead{$M_\odot$} & \colhead{$M_\odot\,\mathrm{yr}^{-1}$} &  \\
\colhead{(1)} & \colhead{(2)} & \colhead{(3)}& \colhead{(4)}& \colhead{(5)}& \colhead{(6)}& \colhead{(7)}& \colhead{(8)} &\colhead{(9)} & \colhead{(10)}
}
\startdata
N-100 & N-40733 & 189.024751 & 62.241858 & 0.734 & p & $41.46^{+0.16}_{-0.21} $ &$8.54^{+0.07}_{-0.08} $ &$-0.60^{+0.05}_{-0.06} $ & AGN \\
N-111 & N-50937 & 189.034350 & 62.264732 & 0.478 & p & $42.18^{+0.04}_{-0.04} $ &$8.46^{+0.31}_{-0.31} $ &$-0.10^{+0.45}_{-0.45} $ & AGN \\
N-213 & N-25982 & 189.129861 & 62.210093 & 0.457 & s & $40.70^{+0.15}_{-0.17} $ &$9.30^{+0.04}_{-0.05} $ &$0.44^{+0.03}_{-0.03} $ & Gal \\
N-253 & N-31232 & 189.151175 & 62.222187 & 0.680 & s & $40.79^{+0.20}_{-0.24} $ &$9.44^{+0.10}_{-0.12} $ &$0.30^{+0.14}_{-0.20} $ & AGN \\
N-289 & N-18818 & 189.173444 & 62.192166 & 0.089 & s & $39.55^{+0.08}_{-0.09} $ &$8.65^{+0.03}_{-0.03} $ &$-0.67^{+0.04}_{-0.04} $ & Gal \\
N-308 & N-21628 & 189.187722 & 62.199480 & 1.603 & p & $41.77^{+0.19}_{-0.24} $ &$8.71^{+0.11}_{-0.15} $ &$0.15^{+0.04}_{-0.05} $ & AGN \\
N-312 & N-47542 & 189.189311 & 62.256444 & 0.682 & s & $41.07^{+0.15}_{-0.17} $ &$9.27^{+0.02}_{-0.02} $ &$0.71^{+0.02}_{-0.02} $ & Gal \\
N-321 & N-16375 & 189.191993 & 62.186450 & 0.292 & p & $39.87^{+0.23}_{-0.29} $ &$7.26^{+0.17}_{-0.28} $ &$-2.92^{+0.33}_{-0.33} $ & AGN \\
N-335 & N-40537 & 189.201260 & 62.240720 & 0.139 & s & $39.69^{+0.11}_{-0.12} $ &$9.46^{+0.02}_{-0.02} $ &$0.13^{+0.02}_{-0.03} $ & Gal \\
N-354 & N-50640 & 189.214019 & 62.264555 & 1.328 & p & $41.74^{+0.16}_{-0.18} $ &$8.62^{+0.06}_{-0.07} $ &$-0.06^{+0.05}_{-0.05} $ & AGN \\
N-359 & N-44432 & 189.217185 & 62.249258 & 0.358 & s & $40.34^{+0.17}_{-0.20} $ &$9.13^{+0.02}_{-0.03} $ &$-0.19^{+0.03}_{-0.04} $ & Gal \\
N-374 & N-13712 & 189.225170 & 62.179521 & 0.595 & s & $40.81^{+0.19}_{-0.24} $ &$9.18^{+0.09}_{-0.11} $ &$0.33^{+0.05}_{-0.06} $ & Gal \\
N-377 & N-49645 & 189.226275 & 62.262160 & 0.440 & p & $40.53^{+0.18}_{-0.21} $ &$7.58^{+0.11}_{-0.16} $ &$-1.74^{+0.08}_{-0.10} $ & AGN \\
N-384 & N-19158 & 189.231780 & 62.193174 & 0.559 & s & $40.70^{+0.19}_{-0.21} $ &$9.13^{+0.06}_{-0.07} $ &$-0.01^{+0.08}_{-0.10} $ & AGN \\
N-396 & N-39498 & 189.238208 & 62.239585 & 1.223 & p & $41.32^{+0.23}_{-0.28} $ &$8.83^{+0.08}_{-0.10} $ &$-0.03^{+0.23}_{-0.52} $ & AGN \\
N-406 & N-50259 & 189.243281 & 62.263530 & 0.457 & s & $40.61^{+0.19}_{-0.23} $ &$9.36^{+0.03}_{-0.03} $ &$0.49^{+0.03}_{-0.03} $ & Gal \\
N-427 & N-17912 & 189.258482 & 62.189687 & 0.136 & s & $39.60^{+0.13}_{-0.15} $ &$9.38^{+0.07}_{-0.08} $ &$-0.20^{+0.14}_{-0.20} $ & Gal \\
N-437 & N-17748 & 189.265449 & 62.189735 & 0.910 & s & $41.15^{+0.20}_{-0.25} $ &$9.22^{+0.05}_{-0.06} $ &$0.35^{+0.07}_{-0.09} $ & AGN \\
N-448 & N-23560 & 189.271613 & 62.204412 & 0.331 & p & $40.00^{+0.24}_{-0.30} $ &$7.21^{+0.15}_{-0.24} $ &$-3.91^{+0.53}_{-0.53} $ & AGN \\
N-541 & N-6502 & 189.344411 & 62.159808 & 0.555 & p & $42.11^{+0.04}_{-0.04} $ &$9.04^{+0.16}_{-0.25} $ &$-6.08^{+0.46}_{-0.46} $ & AGN \\
N-544 & N-14029 & 189.349113 & 62.179417 & 0.113 & s & $39.53^{+0.14}_{-0.16} $ &$9.27^{+0.02}_{-0.02} $ &$-0.30^{+0.09}_{-0.11} $ & Gal \\
S-235 & S-34126 & 53.054061 & -27.810221 & 0.417 & s & $40.15^{+0.18}_{-0.26} $ &$8.72^{+0.06}_{-0.07} $ &$-0.83^{+0.11}_{-0.15} $ & AGN \\
S-257 & S-28151 & 53.059607 & -27.822467 & 0.665 & s & $40.68^{+0.16}_{-0.21} $ &$9.12^{+0.08}_{-0.10} $ &$0.44^{+0.03}_{-0.04} $ & Gal \\
S-277 & S-27209 & 53.063884 & -27.824325 & 0.338 & s & $39.93^{+0.15}_{-0.19} $ &$8.67^{+0.07}_{-0.08} $ &$-0.47^{+0.13}_{-0.18} $ & Gal \\
S-285 & S-38943 & 53.066425 & -27.800698 & 1.351 & p & $41.32^{+0.18}_{-0.22} $ &$8.71^{+0.06}_{-0.07} $ &$-0.42^{+0.24}_{-0.59} $ & AGN \\
S-316 & S-25235 & 53.073240 & -27.828132 & 0.338 & s & $40.02^{+0.14}_{-0.17} $ &$8.43^{+0.03}_{-0.03} $ &$-0.13^{+0.03}_{-0.04} $ & Gal \\
S-339 & S-26815 & 53.078854 & -27.824887 & 0.330 & s & $39.97^{+0.13}_{-0.15} $ &$8.48^{+0.04}_{-0.04} $ &$-0.65^{+0.02}_{-0.02} $ & Gal \\
S-340 & S-39977 & 53.079175 & -27.798800 & 0.666 & s & $42.50^{+0.01}_{-0.01} $ &$9.19^{+0.16}_{-0.27} $ &$0.33^{+0.42}_{-0.42} $ & AGN \\
S-344 & S-35327 & 53.079957 & -27.808786 & 0.862 & p & $41.21^{+0.16}_{-0.20} $ &$9.05^{+0.03}_{-0.03} $ &$0.51^{+0.03}_{-0.03} $ & AGN \\
S-404 & S-2304 & 53.094878 & -27.873374 & 0.131 & s & $40.06^{+0.04}_{-0.05} $ &$8.97^{+0.04}_{-0.04} $ &$-0.59^{+0.11}_{-0.14} $ & Gal \\
S-414 & S-18532 & 53.096647 & -27.841494 & 3.690 & p & $42.87^{+0.20}_{-0.27} $ &$9.21^{+0.06}_{-0.07} $ &$0.66^{+0.07}_{-0.08} $ & AGN \\
S-456 & S-56467 & 53.106962 & -27.765523 & 1.035 & s & $40.94^{+0.19}_{-0.23} $ &$9.45^{+0.10}_{-0.12} $ &$0.77^{+0.11}_{-0.14} $ & Gal \\
S-471 & S-60366 & 53.108588 & -27.757328 & 2.877 & p & $42.49^{+0.15}_{-0.18} $ &$9.14^{+0.10}_{-0.13} $ &$0.45^{+0.07}_{-0.08} $ & AGN \\
S-477 & S-28977 & 53.109596 & -27.820896 & 0.339 & s & $40.44^{+0.09}_{-0.09} $ &$9.32^{+0.02}_{-0.02} $ &$-1.80^{+0.02}_{-0.02} $ & AGN \\
S-487 & S-37757 & 53.111527 & -27.803048 & 0.696 & p & $40.55^{+0.18}_{-0.23} $ &$8.13^{+0.03}_{-0.04} $ &$-2.05^{+0.08}_{-0.09} $ & AGN \\
S-505 & S-56330 & 53.114919 & -27.765741 & 0.309 & s & $39.85^{+0.14}_{-0.16} $ &$7.55^{+0.06}_{-0.07} $ &$-1.58^{+0.03}_{-0.03} $ & AGN \\
S-510 & S-49090 & 53.116106 & -27.780301 & 1.814 & p & $41.73^{+0.15}_{-0.18} $ &$8.49^{+0.09}_{-0.11} $ &$-0.19^{+0.05}_{-0.06} $ & AGN \\
S-540 & S-6260 & 53.122194 & -27.865452 & 0.510 & s & $40.63^{+0.12}_{-0.15} $ &$8.97^{+0.02}_{-0.02} $ &$0.10^{+0.02}_{-0.02} $ & Gal \\
S-575 & S-32347 & 53.128867 & -27.813904 & 2.357 & p & $42.21^{+0.19}_{-0.24} $ &$8.75^{+0.04}_{-0.04} $ &$0.19^{+0.03}_{-0.03} $ & AGN \\
S-618 & S-28195 & 53.138398 & -27.822526 & 0.204 & s & $39.66^{+0.11}_{-0.12} $ &$8.69^{+0.02}_{-0.02} $ &$-6.44^{+1.40}_{-1.40} $ & AGN \\
S-634 & S-16673 & 53.141706 & -27.845364 & 1.364 & s & $41.45^{+0.15}_{-0.19} $ &$8.65^{+0.14}_{-0.20} $ &$-0.03^{+0.05}_{-0.06} $ & AGN \\
S-648 & S-46495 & 53.144746 & -27.785512 & 0.247 & s & $39.65^{+0.15}_{-0.18} $ &$8.50^{+0.02}_{-0.02} $ &$-1.05^{+0.02}_{-0.02} $ & Gal \\
S-664 & S-67861 & 53.148800 & -27.738970 & 0.936 & s & $41.13^{+0.15}_{-0.19} $ &$9.48^{+0.08}_{-0.09} $ &$0.63^{+0.09}_{-0.12} $ & Gal \\
S-674 & S-27862 & 53.149930 & -27.823965 & 0.232 & s & $39.46^{+0.18}_{-0.23} $ &$8.59^{+0.02}_{-0.02} $ &$-0.98^{+0.02}_{-0.02} $ & Gal \\
S-692 & S-18882 & 53.153117 & -27.840614 & 3.330 & p & $42.53^{+0.17}_{-0.21} $ &$8.66^{+0.06}_{-0.07} $ &$0.11^{+0.03}_{-0.04} $ & AGN \\
S-695 & S-42775 & 53.154638 & -27.793304 & 0.232 & s & $39.78^{+0.16}_{-0.19} $ &$8.19^{+0.03}_{-0.03} $ &$-0.94^{+0.02}_{-0.02} $ & AGN \\
S-705 & S-8692 & 53.156232 & -27.860848 & 0.687 & s & $41.02^{+0.12}_{-0.15} $ &$8.61^{+0.05}_{-0.06} $ &$-0.07^{+0.05}_{-0.05} $ & AGN \\
S-720 & S-6376 & 53.160860 & -27.865316 & 0.416 & s & $40.34^{+0.15}_{-0.20} $ &$9.28^{+0.09}_{-0.11} $ &$-0.27^{+0.10}_{-0.12} $ & Gal \\
S-763 & S-6450 & 53.170877 & -27.864778 & 0.233 & p & $40.12^{+0.08}_{-0.10} $ &$6.88^{+0.12}_{-0.17} $ &$-2.68^{+0.12}_{-0.16} $ & AGN \\
S-768 & S-59833 & 53.172973 & -27.759118 & 0.147 & s & $39.74^{+0.07}_{-0.08} $ &$8.95^{+0.02}_{-0.02} $ &$-2.18^{+0.02}_{-0.02} $ & Gal \\
S-778 & S-40427 & 53.174770 & -27.799341 & 0.151 & s & $39.43^{+0.14}_{-0.16} $ &$8.91^{+0.02}_{-0.02} $ &$-0.66^{+0.02}_{-0.02} $ & Gal \\
S-793 & S-17383 & 53.179699 & -27.843940 & 0.734 & s & $40.88^{+0.17}_{-0.24} $ &$9.25^{+0.15}_{-0.23} $ &$0.57^{+0.05}_{-0.05} $ & Gal \\
S-828 & S-23105 & 53.189312 & -27.832328 & 1.530 & p & $41.56^{+0.17}_{-0.24} $ &$9.12^{+0.11}_{-0.14} $ &$0.57^{+0.16}_{-0.26} $ & AGN \\
S-857 & S-31759 & 53.199941 & -27.815534 & 0.233 & s & $39.91^{+0.13}_{-0.15} $ &$9.27^{+0.02}_{-0.02} $ &$-0.31^{+0.05}_{-0.06} $ & Gal \\
S-860 & S-57431 & 53.201806 & -27.764137 & 0.214 & s & $39.82^{+0.15}_{-0.19} $ &$8.81^{+0.02}_{-0.02} $ &$-0.75^{+0.02}_{-0.02} $ & Gal \\
S-862 & S-59859 & 53.203517 & -27.758832 & 0.276 & s & $40.47^{+0.13}_{-0.18} $ &$9.18^{+0.02}_{-0.02} $ &$-1.95^{+0.22}_{-0.47} $ & AGN \\
\hline
\enddata
\tablecomments{This table is sorted in ascending order as (1) XID, the X-ray source ID in \citet{Xue2016ApJS..224...15X} and \citet{Luo2017ApJS..228....2L}. (2) DJA source ID. (3) and (4) J2000 coordinates. (5) Redshift. (6) Redshift type. ``s" and ``p" indicate spectroscopic and photometric redshift, respectively. (7) Observed 2--10 keV luminosity. (8) and (9) \texttt{CIGALE} best-derived $M_\ast$ and SFR with AGN template. (10) AGN classification based on Section~\ref{subsec:agn_classification}.}
\end{deluxetable}

\section{AGNs in Extreme Dwarf Galaxies}\label{app:edg-agn}
\begin{figure}
    \begin{minipage}[c]{0.56\linewidth}
        \centering
        \includegraphics[width=\linewidth]{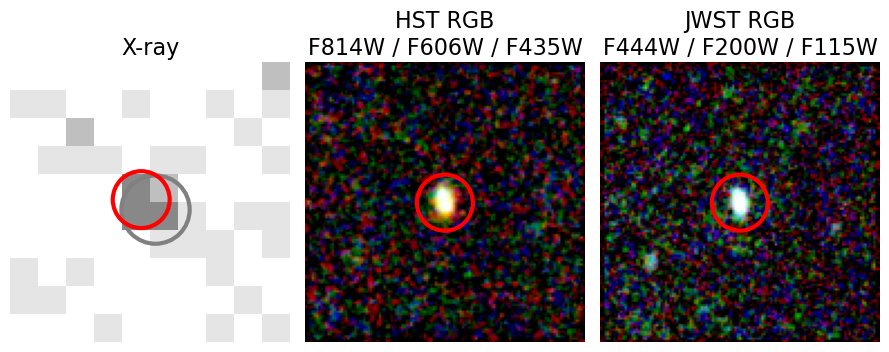}
    \end{minipage}%
    \hfill
    \begin{minipage}[c]{0.42\linewidth}
        \centering
        \includegraphics[width=\linewidth]{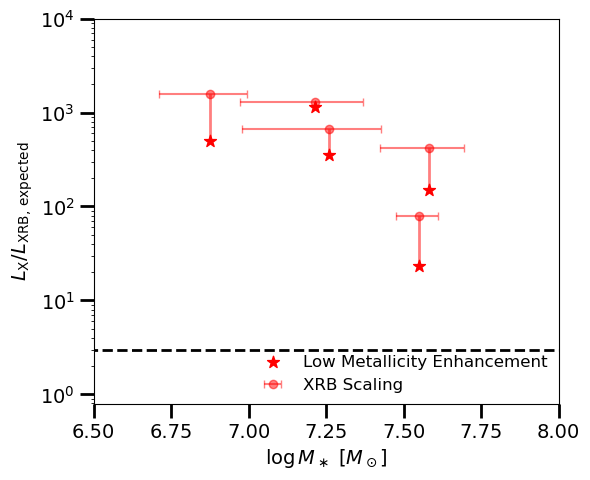}
    \end{minipage}
    \caption{
    \textit{Left:} Multiwavelength cutouts of the representative EDG source N-377. From left to right, the panels show the X-ray image, an HST color composite constructed from F814W/F606W/F435W, and a JWST color composite constructed from F444W/F200W/F115W. The red circle marks the optical--NIR counterpart, while the gray circle indicates the X-ray positional uncertainty. The host galaxy is compact and does not exhibit an obvious distinct nuclear point source in either the HST or JWST imaging.
    \textit{Right:} Ratio of the observed 2--10~keV X-ray luminosity to the expected XRB luminosity as a function of stellar mass for the five EDG AGN candidates. Red circles show the ratios calculated using the fiducial XRB scaling relation adopted in the main analysis, while red stars show the ratios after applying the low-metallicity HMXB enhancement from \citet{Lehmer2022ApJ...930..135L}. Vertical line segments connect the fiducial and metallicity-enhanced estimates for each source, and the horizontal error bars indicate the uncertainties in stellar mass. The dashed line marks the adopted X-ray-excess threshold of $L_\mathrm{X}/L_\mathrm{XRB,\,expected}=3$. Even after accounting for enhanced HMXB emission, all five EDG AGN candidates remain substantially more X-ray luminous than expected from their stellar populations.}
    \label{fig:edg}
\end{figure}

In this work, we identify five AGN candidates in extreme dwarf galaxies (EDGs)  with derived stellar masses of $M_\ast < 10^{8}~M_\odot$: N-321, N-377, N-448, S-505, and S-763 (Table~\ref{tab:catalog}). The left panel of Figure~\ref{fig:edg} shows N-377 as a representative example, while N-377, N-448, and S-505 are also presented in Figure~\ref{fig:grid_image}. 
These sources represent the lowest-mass systems in our X-ray-selected dwarf-AGN sample and therefore warrant additional consideration, because the X-ray emission of such low-mass galaxies can be especially sensitive to recent star formation, metallicity-dependent HMXB production, and stochastic sampling of the XRB population. 

The classification of these systems as EDGs depends on their SED-derived stellar masses, which remain subject to systematic uncertainties associated with bursty star-formation histories and/or rapidly varying stellar mass-to-light ratios \citep[e.g.,][]{Dominguez2015MNRAS.451..839D,Guo2016ApJ...833...37G}. For our sources, however, both the SED fits and the image-decomposition analysis indicate that the rest-frame optical and NIR emission is dominated by the host galaxy rather than by an unresolved AGN component (Sections~\ref{sec:sed_fitting_dwarf_selection} and \ref{sec:nuclear_ps}). AGN contamination is therefore unlikely to be the dominant source of uncertainty in the inferred stellar masses. In addition, our SED modeling includes nebular emission, reducing potential biases from strong emission lines. Nevertheless, uncertainties in the adopted star-formation histories and stellar mass-to-light ratios could shift individual objects across the nominal $M_\ast=10^{8}~M_\odot$ boundary. We therefore refer to them as EDG AGN candidates. 

All five EDG candidates exhibit X-ray luminosities above the expected stellar contribution from XRBs and hot gas, based on the scaling relations adopted in the main analysis. 
However, low-metallicity star-forming galaxies can show enhanced HMXB emission relative to solar-metallicity systems, particularly following recent star-formation episodes \citep[e.g.,][]{Fragos2013ApJ...764...41F,Lehmer2016ApJ...825....7L,Lehmer2021ApJ...907...17L}. 
This effect is especially relevant for EDGs, whose low stellar masses and potentially metal-poor stellar populations may increase the expected X-ray output per unit star formation.

To assess whether the 2--10 keV X-ray luminosities of these EDG candidates could be explained by enhanced stellar processes alone, we repeat the XRB comparison using the metallicity-dependent HMXB prescription of \citet{Lehmer2022ApJ...930..135L}. 
This prescription increases the predicted XRB luminosities, as shown in the right panel of Figure~\ref{fig:edg}. 
Nevertheless, all five sources remain above the adopted X-ray-excess threshold (dashed line) after accounting for low-metallicity HMXB enhancement. 
Thus, while enhanced HMXB emission may reduce the inferred level of X-ray excess, it does not fully account for the observed X-ray output of the EDG AGN candidates.

X-ray AGN candidates in galaxies with $M_\ast<10^{8}~M_\odot$ remain rare and poorly studied. 
These systems are therefore valuable targets for future follow-up observations. 
In particular, JWST spectroscopy could test if these sources host accreting massive black holes by improving nebular emission-line diagnostics and searching for narrow or broad high-ionization lines. Its high spatial resolution could help isolate compact nuclear emission from that of the low-mass host galaxy. 
Such observations would also help distinguish weak accretion signatures from stellar processes in the extreme low-mass regime.

\bibliography{dwarfAGN}{}
\bibliographystyle{aasjournalv7}


\end{CJK*}
\end{document}